\documentclass[11pt]{article}
\usepackage{amssymb,jheppub,slashed}
\usepackage{braket,commath,nicefrac}
\usepackage{subcaption,siunitx,amsmath}
\usepackage{slashed,float}
\usepackage{xcolor}
\usepackage{braket}
\usepackage{scalerel}
\usepackage{epsfig}
\usepackage[utf8]{inputenc}
\usepackage{color}
\usepackage{dsfont} 
\usepackage{amsbsy} 
\usepackage{mathrsfs} 
\usepackage{graphicx}
\usepackage{hyperref}

\newcommand{\nn}{\nonumber}
\newcommand{\be}{\begin{equation}}
\newcommand{\ee}{\end{equation}}
\newcommand{\bea}{\begin{eqnarray}}
\newcommand{\eea}{\end{eqnarray}}
\newcommand{\ba}{\begin{array}}
\newcommand{\ea}{\end{array}}
\newcommand{\bi}{\begin{itemize}}
\newcommand{\ei}{\end{itemize}}

\renewcommand{\vec}[1]{\mbox{\boldmath $#1 \!\!$ \unboldmath}}

\newcommand{\lf}{\left}
\newcommand{\rg}{\right}

\title{Controlling Quantum discord and steering in Electron-Positron Annihilation Using Polarized Beams}
\author[a,b]{Hong-Wei Zhang,}
\author[b,c,d]{Xu Cao,}
\author[a]{Tai-Fu Feng}

\affiliation[a]{Department of Physics, Hebei University, Baoding 071002, China}
\affiliation[b]{State Key Laboratory of Heavy Ion Science and Technology, Institute of Modern Physics, Chinese Academy of Sciences, Lanzhou 730000, China}
\affiliation[c]{University of Chinese Academy of Sciences, Beijing 100049, China}
\affiliation[d]{Research Center for Hadron and CSR Physics, Lanzhou University and Institute of Modern Physics of CAS, Lanzhou 730000, China}

\emailAdd{hweiz0929@163.com}
\emailAdd{caoxu@impcas.ac.cn}
\emailAdd{fengtf@hbu.edu.cn}

\date{\today}

\abstract{
  Quantum discord and steering offer crucial insights into the non-classical nature of hyperon-antihyperon pairs, a massive two-qubit system produced in high energy electron-positron annihilation.
  This work theoretically investigates the generation and control of these quantum correlations by leveraging longitudinal and transverse polarization of lepton beams.
  By exploiting the joint spin density matrix of hyperon pairs, the sensitivity of quantum discord and steering to the beam polarization degree are numerically quantified.
  Our analysis reveals distinct angular regimes where beam polarization can enhance steering and discord.
  Hierarchy of different quantum correlations are examined under the case of  polarized lepton beams by constructing a measure in the spirit of entanglement of formation.
  It is confirmed that quantum discord remain non-zero even in regions with vanishing entanglement corresponding to separable states, controlled via transversely polarized beams.
  As an experimentally tunable parameter, beam polarization offers an effective means to manipulate the quantum correlation of hyperon-antihyperon systems, thereby providing a practical route for preparing and probing quantum states in high-energy particle physics.
}

\begin{document}

\maketitle

\section{Introduction}

Studies of entangled states in low-energy photonic and atomic systems have probed the quantum nonlocal correlations, the defining characteristic of quantum mechanics \cite{Aspect:1981zz,Wiseman:2007hyt,Giustina:2015yza}.
Entanglement associated with quantum information at hadron and lepton colliders continues to be an active area of investigation \cite{Barr:2024djo,Afik:2025ejh}.
The top quark, owing to its unique property of decaying before hadronization, retains its spin information, making it an ideal system for exploring quantum entanglement and coherence \cite{Afik:2020onf,Afik:2022kwm,Varma:2023gwh,Rai:2025qke,Aguilar-Saavedra:2023hss,Aguilar-Saavedra:2024hwd,Arai:2026jtc,Subba:2026nzs}.
A variety of other collider processes involving photons \cite{Cheng:2026ktp,deLima:2026jbb}, $\tau$ and $\mu$ leptons~\cite{Ehataht:2023zzt,Ma:2023yvd,LoChiatto:2024dmx,Aguilar-Saavedra:2023lwb}, neutrinos \cite{Formaggio:2016cuh,Bittencourt:2023asd}, massive gauge bosons~\cite{Aguilar-Saavedra:2022wam},
hyperons~\cite{Tornqvist:1980af,Hao:2009kj,Wu:2024asu,Wu:2024mtj,Hong:2025drg,Jaloum:2025bkx,Wu:2025dds,Pei:2026rlh,Pei:2025ito,Fabbrichesi:2024rec,Pei:2025yvr,Hong:2025drg,Lin:2025eci,Wang:2026nls} and different quark flavors~\cite{Cheng:2025cuv,Afik:2025grr,Fucilla:2025kit,Qi:2025onf,Cheng:2025zaw,Hatta:2025obw,Choi:2026omc,Liu:2026dzv} has been an ongoing concern.
The loopholes are carefully scrutinized, addressing whether a Bell test of local realism is in principle feasible at high energy colliders~\cite{Abel:1992kz,Dreiner:1992gt,Li:2024luk,Low:2025aqq,Bechtle:2025ugc,Abel:2025skj,Ai:2025wnt,Pei:2026wfu,Aguilar-Saavedra:2026rsx}.
Experimental tests of local realism with entangled top-quark pairs at LHCb \cite{ATLAS:2023fsd,CMS:2024pts} and hyperon pairs at BEPCII \cite{BESIII:2025vsr} continue to pose significant challenges and opportunities in high-energy particle physics.
On the other hand, quantum information offers new insights into the strong force through alternative perspectives, such as entanglement suppression \cite{Beane:2018oxh} and maximal entanglement \cite{Cervera-Lierta:2017tdt,Kharzeev:2017qzs}.
Quantum entanglement between partons inside hadron is exploited to give different insight into the hadron structure \cite{Fucilla:2026mkg,Qian:2026aoh,Zhang:2025ean,Dumitru:2025bib,Qian:2024fqf,Bloss:2026yrf,Agrawal:2026zwa}.

Investigating quantum tomography in high-energy physics involves two new scenarios of entanglement manipulation.
One arises when particles decay, potentially leading to differences in entanglement between the decay products and the parent fermion pairs' system \cite{Aguilar-Saavedra:2024fig,Feng:2025ryr,Li:2026bkf,Chen:2026oaf}.
The other occurs when beam polarization is used as a tool to tune the entanglement of the produced particle pairs \cite{Guo:2026yhz,Altakach:2026fpl,Zhang:2026nwm,Fang:2026ddi}.
To characterize the entanglement resource of the mixed states of two qubits such as hyperon-antihyperon pairs, concurrence~\cite{Wootters:1997id}
and negativity~\cite{Vidal:2002zz} are two practical and widely
used entanglement measures conceptualized within the entanglement-separability paradigm.
The former is based on entanglement of formation, and latter is derived from the positive partial transpose criterion (PPT)
criterion for separability~\cite{Horodecki:1996nc}.
Our previous work investigated how longitudinal and transverse polarization of lepton beams manipulate the concurrence, negativity and Bell nonlocality of hyperon pair system \cite{Zhang:2026nwm}.

Long before the advent of concurrence and negativity, the quantum steering was first introduced by Schr\"{o}dinger \cite{Schrodinger:2008pyl} in response to the Einstein-Podolsky-Rosen (EPR) paradox \cite{Einstein:1935rr}.
However, only after a systematic criteria in the language of quantum
information processing was developed \cite{Reid:1989zvj,Wiseman:2007hyt}, steering receives much attention.
Quantum steering refers to the nonlocal ability of local measurements
performed on one subsystem of an entangled particle pair to influence the quantum state of another subsystem,
even when the subsystems are spatially separated \cite{Bera:2017lmd,Uola:2020kps}.
Steerability is regarded as a novel form of inseparability in quantum mechanics, intermediate between entanglement and nonlocality.
Steerability is an inherently asymmetric correlation, and there are states where one of states can steer another but not the other way round.

As another asymmetric form of quantum correlations, quantum discord is the most basic form of quantum correlations, more general than entanglement, in the sense that quantum states of two or more parties that are not entangled, and yet quantum correlated \cite{Ollivier:2001fdq,Henderson:2001wrr}.
Discord is not only a signature of quantumness rather it is also a signature of nonclassicality, capturing how much a physical system’s
state deviates from classical descriptions.
Experiments in nuclear and optical system have demonstrated the
existence of quantum discord and related quantities \cite{Bennett:1998ev}.
Quantum discord and steering have been proposed as tools to investigate quantum information properties in top-quark pairs at LHC \cite{Afik:2022dgh}.
Compact closed-form expressions for the quantum discord of top-antitop quarks pairs are provided, enabling efficient computation across the entire phase space \cite{Han:2024ugl}.
Unlike elementary particle systems, it is found that hyperon–antihyperon systems exhibit distinct characteristics in their quantum discord and steering \cite{Wu:2025dds}, and the significance of quantum coherence is further underscored \cite{Jaloum:2025bkx}.
In this work, we investigate hyperon–antihyperon pairs produced in polarized electron–positron collisions by analyzing the dynamics of quantum correlations, specifically through discord and steering. Our results reveal how these correlations respond uniquely to the polarization of lepton beams, thereby providing a comprehensive description of the quantum information landscape for this process.

This paper is organized as follows.
In Sec.~\ref{sec:preparation}, the formalism of quantum discord and steering are introduced based on the polarization vector, spin correlation matrix and eigenvalue decomposition incorporating lepton beam polarization.
The influence of longitudinal and transverse beam polarization is present in Sec.~\ref{sec:PL} and Sec.~\ref{sec:PT}, respectively.
Hierarchy of quantum correlations, quantum discord under transverse beam polarization, {and experimental sensitivity} are discussed in Sec.~\ref{sec:discussion}, and the conclusion is briefly given in Sec.~\ref{sec:conclusion}.

\section{Formalism} \label{sec:preparation}

\subsection{Bloch--Fano Coefficient} \label{sec:bfc}

A pair of octet hyperon-antihyperon ($Y\bar{Y}$) can be considered as a massive $2$-qubit system, which can be fully described by a spin density matrix $\rho_{Y\bar{Y}}$ in Bloch-Fano representation\cite{Fano:1983zz}:
\begin{equation}
    \rho_{Y\bar{Y}}= \frac{1}{4} \left(\mathbb{I}+\vec{B}^{+}\cdot\vec{\sigma}\otimes\mathbb{I}+\mathbb{I}\otimes\vec{B}^{-}\cdot\vec{\sigma} +\sum_{i,j=1}^{3} C_{i j}\, {\sigma}_i\otimes {\sigma}_j \right)\label{eq:two_qubit},
\end{equation}
with $\boldsymbol{\sigma} = (\sigma_x,\sigma_y,\sigma_z)$ being the Pauli matrices.
Here $\vec{B}^+=\mathrm{Tr}(\rho_{Y\bar{Y}}\,\vec{\sigma}\otimes\mathbb{I})$ and $\vec{B}^-=\mathrm{Tr}(\rho_{Y\bar{Y}}\,\mathbb{I}\otimes\vec{\sigma})$ are polarization or Bloch vectors of hyperon and antihyperon, respectively.
$\left(C_{ij}\right) =\left( \mathrm{Tr}(\rho_{Y\bar{Y}}\,\sigma_i\otimes\sigma_j) \right)$ is a $3\times3$ spin correlation matrix.

\begin{figure}[!t]
    \centering
    \includegraphics[scale=0.6]{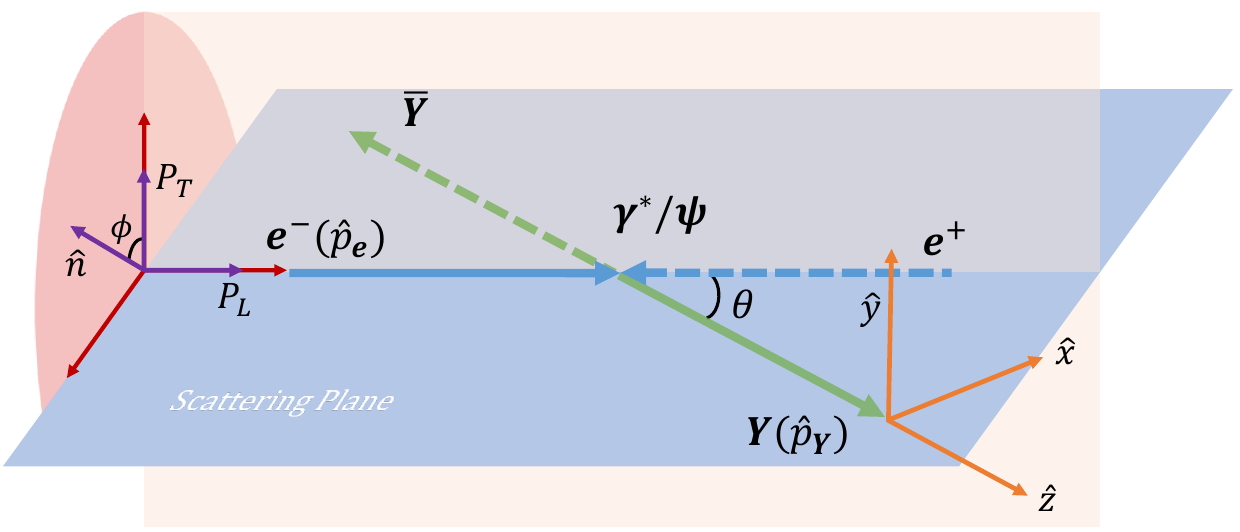}
    \caption{{The coordinate system used for $e^{+}e^{-} \to \gamma^*/\psi \to Y\bar{Y}$ and helicity frame defined for hyperon. The $\hat{\mathbf{z}}$ axis of helcity frame is defined along $\hat{\mathbf{p}}_Y$, representing the hyperon momentum direction in the scattering plane.
    The remaining axes are $\hat{\mathbf{y}} = (\hat{\mathbf{p}}_{e}\times\hat{\mathbf{p}}_{Y}) / \left|\hat{\mathbf{p}}_{e}\times\hat{\mathbf{p}}_{Y}\right|$ with $\hat{\mathbf{p}}_{e}$ being the flight direction of the incoming positron, and $\hat{\mathbf{x}} = \hat{\mathbf{y}}\times\hat{\mathbf{z}}$.}}
    \label{fig:frame}
\end{figure}

The density matrix $\rho_{Y\bar{Y}}$ is frame-dependent.
The center-of-mass (c.m.) frame for the $e^{+}e^{-} \to \gamma^*/\psi \to Y\bar{Y}$ process, as defined in the scattering plane, is illustrated in Fig.~\ref{fig:frame}.
{The $\hat{\mathbf{z}}$ axis of the helicity frame of hyperon is defined by the hyperon's momentum $\hat{\mathbf{p}}_Y$ in the scattering plane and subsequently serves as the spin-quantization axis after boosting to the hyperon rest frame.
Since this boost is performed along the momentum direction, the helicity remains invariant during the transformation \cite{Perotti:2018wxm}.
The $\hat{\mathbf{y}}$ axis is chosen to be perpendicular to the scattering plane.
The $\hat{\mathbf{x}}$ axis is chosen such that $\hat{\mathbf{x}}$, $\hat{\mathbf{y}}$ and $\hat{\mathbf{z}}$ adhere to the right-hand rule.
The helicity frame of antihyperon $\bar{Y}$ is chosen as $\{\hat{\mathbf{x}}_{\bar{Y}},\hat{\mathbf{y}}_{\bar{Y}},\hat{\mathbf{z}}_{\bar{Y}}\}=\{\hat{\mathbf{x}}_Y,\hat{\mathbf{y}}_Y,\hat{\mathbf{z}}_Y\}$, leading to $\vec{B}^{+}=\vec{B}^{-}$ and $\left(C_{ij}\right)$ matrix being real symmetric if $CP$ violation is not considered.}
The calculated quantum discord and steering are independent of this axis convention.

{
Neglecting the electron mass and assuming parity and charge-parity ($CP$) conservation, this $2 \to 2$ fermion scattering process is described by only two independent helicity amplitudes at the leading order of one-photon exchange. In other words, the most general amplitude for this process consists of two independent Lorentz structures, which correspond to the widely used Sachs' electric and magnetic form factors, $G_E$ and $G_M$, respectively. In the timelike region, these form factors are complex due to strong interaction rescattering between the hyperon-antihyperon pairs~\cite{Lin:2021xrc}. Their relative magnitude and phase, typically defined in the literature as $R_{EM} = |G_E/G_M|$ and $\Delta\Phi$, can be experimentally determined through the full reconstruction of the sequential hyperon decays \cite{Perotti:2018wxm}. This contrasts with cases involving elementary particles in the final states, where the rescattering is of a higher order in QED, resulting in a negligible imaginary part. Thus, a non-vanishing $\Delta\Phi$ directly reflects the internal structure of the hyperon.
The parameter $\alpha_{\psi}$, which characterizes the angular distribution of the hyperon in the c.m. frame, is related to $R_{EM}$ via
\begin{equation}
    \alpha_{\psi} = \frac{\tau - R_{EM}^2}{\tau + R_{EM}^2},
\end{equation}
where $\tau = s/4M^2$ and $s$ is the center-of-mass (c.m) energy squared. From a partial-wave perspective, a non-zero $\alpha_\psi$ signifies a $D$-wave contribution in addition to the dominant $S$-wave component~\cite{Wu:2021yfv}.
The parameters $\alpha_\psi$ and $\Delta\Phi$ characterize the intrinsic decay dynamics of the $\gamma^* /\psi \to Y\bar{Y}$ transition. Since they are fully accounted for by the transition amplitude of the virtual photon or charmonium conversion to the hyperon-antihyperon pair, they are independent of how the $\gamma^* /\psi$ is produced, regardless of the initial beam polarization. }
The parameters of $ J/\psi\rightarrow Y\bar{Y}$ measured by BESIII collaboration are listed in Table~\ref{tab:decay_parameters}.

When the electron beam is longitudinally polarized, the Bloch vectors $\vec{B}_L$ and spin correlation matrix $\vec{C}_L$ of density matrix $\rho^{P_L}_{Y\bar{Y}}$ under the $\sigma_z\otimes\sigma_z$ representation are given by \cite{Perotti:2018wxm,Batozskaya:2023rek,Salone:2022lpt,Zeng:2023wqw,Zhang:2024rbl,Zhao:2025cbd}
\begin{align}
&\vec{B}_{L}=\frac{1}{\chi_L}\begin{pmatrix}
P_L\gamma_\psi\sin\theta\\-\beta_{\psi}\sin\theta\cos\theta\\-P_L(1+\alpha_{\psi})\cos\theta
\end{pmatrix},\label{eq:BL}\\
&\vec{C}_L=\frac{1}{\chi_L}
\begin{pmatrix}
\sin^2\theta & 0 &-\gamma_{\psi}\cos{\theta}\sin{\theta}\\
 0 & -\alpha_{\psi}\sin^2\theta & P_L\beta_{\psi}\sin\theta\\
-\gamma_{\psi}\cos{\theta}\sin{\theta} & P_L\beta_{\psi}\sin\theta & \alpha_{\psi}+ \cos^2\theta
\end{pmatrix}, \label{eq:CL}
\end{align}
with $\chi_L =1+\alpha_{\psi}\cos^{2}\theta$, and $\beta_{\psi}=\sqrt{1-{\alpha_{\psi}}^{2}}\sin\Delta\Phi, \; \gamma_{\psi}=\sqrt{1-{\alpha_{\psi}}^{2}}\cos\Delta\Phi$.
Here $\theta$ is the scattering angle between the momenta of incoming electron and outgoing hyperon.
$P_L$ is the longitudinal polarization degree of the electron beam.
{While beam polarization does not modify the underlying decay amplitudes; rather, it serves as a powerful tool to disentangle the dynamical information already encoded in these amplitudes.
Specifically, introducing beam polarization allows us to access new experimental observables that are otherwise averaged out by summing over the initial spins in the unpolarized case. Consequently, polarized electron beams facilitate the investigation of $CP$ violation in hyperon decays, as explored in the literature~\cite{Salone:2022lpt,Cao:2024tvz}, as well as the study of quantum entanglement in hyperon pairs presented herein and in our previous work~\cite{Zhang:2026nwm}. As shown in Eq.~\eqref{eq:BL}, longitudinal beam polarization induces a new transverse polarization component in the final-state hyperon and a new longitudinal polarization component. The first component provides an independent observable to measure $\Delta\Phi$ in future polarized colliders.}

When both electron and positron beams are transversely polarized, the $\vec{B}_{T}$ and the corresponding correlation matrix $\vec{C}_T$ of the density matrix $\rho^{P_T}_{Y\bar{Y}}$ are \cite{Perotti:2018wxm,Batozskaya:2023rek,Salone:2022lpt,Cao:2024tvz}
\begin{align}
\vec{B}_{T}&=\frac{1}{\chi_T}\begin{pmatrix}
-{P_T^2}\beta_{\psi}\sin\theta\sin 2 \phi \\
\beta_{\psi}\sin\theta\cos\theta({P_T^2}\cos 2 \phi-1)\\0
\end{pmatrix}, \label{eq:BT}\\
\notag
\vec{C}_T&=\frac{1}{\chi_T}
\begin{pmatrix}
\sin^2\theta &0 & -\gamma_{\psi}\sin\theta\cos\theta\\
0&-\alpha_{\psi}\sin^2\theta & 0\\
 -\gamma_{\psi}\sin\theta\cos\theta& 0&\alpha_{\psi}+\cos^2\theta
\end{pmatrix}\\
&+\frac{P_T^2}{\chi_T}
\begin{pmatrix}
(\alpha_{\psi}+\cos^2\theta)\cos2\phi & (1+\alpha_{\psi})\cos\theta\sin 2 \phi &\gamma_{\psi}\sin\theta\cos\theta\cos 2 \phi\\
(1+\alpha_{\psi})\cos\theta\sin 2 \phi&-(1+\alpha_{\psi}\cos^2\theta)\cos 2 \phi &\gamma_{\psi}\sin\theta\sin 2 \phi\\
\gamma_{\psi}\sin\theta\cos\theta\cos 2 \phi& \gamma_{\psi}\sin\theta\sin 2 \phi&\sin^2\theta\cos 2 \phi
\end{pmatrix},
\label{eq:CT}
\end{align}
with $\chi_T =1+\alpha_{\psi}\cos^{2}\theta+{P_T^2}\alpha_{\psi}\sin^2\theta\cos 2\phi$.
Where $\phi$ is the azimuthal angle of hyperons with respect to scattering plane.
$P_T$ is the transverse polarization degree of lepton beams.
Here both lepton beams are assumed to have an equal degree of transverse polarization with their polarization directions aligned in anti-parallel \cite{Cao:2024tvz}.
The transverse polarization of beams induces the azimuthal angle $\phi$ dependence of polarization and spin correlation of produced hyperons, {providing new observables for the determination of $\alpha_\psi$ and $\Delta\Phi$}.
It is noted that for certain combinations of azimuthal angle $\phi$ and polarization degrees $P_T$, the hyperon polarization vanishes completely.

\begin{table}
  \begin{center}
  \resizebox{\linewidth}{!}{
    \begin{tabular}{lccc}
\hline
Decay channel & $\alpha_{\psi}$ & $\Delta\Phi/\mathrm{rad}$ & Ref. \\
\hline
$J/\psi\rightarrow\Lambda\bar{\Lambda}$ & $0.4748 \pm 0.0022 \pm0.0031$ & $0.748\pm0.006\pm0.004$ & \cite{BESIII:2025wxe,BESIII:2022qax,BESIII:2018cnd,BESIII:2017kqw}\\
$J/\psi\rightarrow\Sigma^{+}\bar{\Sigma}^{-}$ & $-0.5047\pm0.0018\pm0.0010$ & $-0.2744\pm0.0033\pm0.0010$ & \cite{BESIII:2025jxt,BES:2008hwe,BESIII:2020fqg}\\
$J/\psi\rightarrow\Sigma^{0}\bar{\Sigma}^{0}$ & $-0.4133\pm0.0035\pm0.0077$ & $-0.0828\pm0.0068\pm0.0033$ & \cite{BESIII:2024nif}\\
$J/\psi\rightarrow\Xi^{-}\bar{\Xi}^{+}$ & $ 0.5851\pm0.0044\pm0.0034$ & $1.2205\pm0.0159\pm0.0056$ & \cite{BESIII:2026hgj,BESIII:2021ypr}\\
$J/\psi\rightarrow\Xi^{0}\bar{\Xi}^{0}$ & $0.514\pm0.006\pm0.015$ & $1.168\pm0.019\pm0.018$ & \cite{BESIII:2016nix,BESIII:2023drj}\\
\hline
    \end{tabular}
    }
  \end{center}
    \caption{The parameters of $J/\psi\rightarrow Y\bar{Y}$ decay
for $Y = \Lambda$, $\Sigma^{+}$, $\Sigma^{0}$, $\Xi^{-}$ and $\Xi^{0}$ measured by BESIII collaboration.}
\label{tab:decay_parameters}
\end{table}

Any density matrix as a Hermitian operator can always be diagonalized into the form $\rho = \sum_{i} \lambda_i |i\rangle\langle i|$, where $\lambda_i$ are the eigenvalues and $\ket{i}$ are the corresponding eigenstates, satisfying $\langle \lambda_{i}|\lambda_{j}\rangle = \delta_{ij}$ and $\sum_{i}\lambda_i = 1$.
For the cases of longitudinal or transverse beams polarization,
the non-zero eigenvalues $\lambda^{L,T}$ of $\rho_{Y\bar{Y}}$ are given by
\begin{align}
    &\lambda^L_{1,2}=\frac{1}{2}\left(1\mp\frac{1}{\chi_L}{\sqrt{\alpha_\psi^2 \sin^4\theta + P_L^2\eta}}\right),\label{eq:eigenPL}\\
    &\lambda^T_{1,2}=\frac{1}{2}\left(1\mp\frac{1}{\chi_T}{\sqrt{\chi_T^2-(1-P_T^4)\eta}}\right),\label{eq:eigenPT}
\end{align}
with $\eta=(1+{\alpha_\psi})(1+{\alpha_\psi \cos2\theta})$.
Note that each case has two non-zero eigenvalues, e.g. the $\rho_{Y\bar{Y}}$ are rank-$2$ matrices under the cases of both unpolarized and polarized lepton beams.
The underlying reason is that polarization of lepton beams does not affect the rank of the spin density matrix of vector charmonium (or virtual photon) of rank-$2$ \cite{Cao:2024tvz}:
\be
\rho^{\gamma^*/\psi} = \frac{1}{2}
\begin{pmatrix}
1 + P_L & 0& P_T^2 \\
0&0&0\\
P_T^2 &0& 1 - P_L \\
\end{pmatrix}.\label{eq:ms}
\ee
See Appendix.~\ref{sec:eigen} for the eigen-states of $\rho_{Y\bar{Y}}^{P_L}$.
The eigen-states of $\rho_{Y\bar{Y}}^{P_T}$ is too lengthy and is therefore omitted for brevity.

For a 2-qubit system in Eq.~\eqref{eq:two_qubit}, the marginal density matrix $\rho_{Y(\bar{Y})}$ is given by \cite{Fano:1983zz}
\begin{equation}
  \rho_{Y(\bar{Y})}=\mathrm{Tr}_{\bar{Y}(Y)}(\rho_{Y(\bar{Y})})=\frac{1}{2}\left(\mathbb{I}+\vec{B}^{\pm}\cdot\vec{\sigma}\right).\label{eq:marginal}
\end{equation}
The eigenvalues $\mu_{1,2}$ of $\rho_{Y(\bar{Y})}$ are uniquely determined by the corresponding polarization vector
\begin{equation}
  \mu_{1,2}=\frac{1}{2}\left(1\mp\left\|\vec{B}^{\pm}\right\| \right),\label{eq:mariginal eigen}
\end{equation}
where $\left\|\vec{B}\right\|=\sqrt{\mathrm{Tr}(\vec{B}^{T}\vec{B})}$ are the trace norm of $\vec{B}$, which can be read from Eqs.~(\ref{eq:BL},\ref{eq:BT}):
\begin{align}
    &\left\|\vec{B}_{L}^{\pm}\right\| =\frac{1}{\chi_L}{\sqrt{\beta_\psi ^2\sin^2\theta\cos^2\theta +P_L^2 \left(\beta_{\psi}^2\cos^2 \theta+\gamma_\psi^2\right)}},\label{eq:rdmeigenPL}\\
    &\left\|\vec{B}_{T}^{\pm}\right\| =\frac{\beta_\psi \sin\theta}{\chi_T}\sqrt{\left(1-P_T^2 \cos 2\phi \right)^2\cos^2\theta + P_T^4 \sin^2 2\phi }.\label{eq:rdmeigenPT}
\end{align}
for the longitudinal and transverse polarization of beams, respectively.
The case of $|\alpha_\psi| = 1$ corresponds to elementary particle pairs, such as top quarks, as well as $\tau$ and $\mu$ leptons, with vanishing polarization $\|\vec{B}^{\pm}\| = 0$ \cite{Guo:2026yhz,Altakach:2026fpl,Fang:2026ddi}.

The state of the hyperon pair can be expressed as a convex combination of density matrices within the spin-triplet subspace, given an appropriate choice of coordinate system~\cite{Wu:2024asu}.
The absence of the spin-singlet state is a direct consequence of angular momentum conservation in the decay of the spin-1 $J/\psi$ (or virtual photon)~\citep{Haidenbauer:2016won,Dai:2023vsw}.
This can be verified using the symmetrized operator $\vec{\sigma}\cdot\vec{\sigma}=\sum_i \sigma_i \otimes \sigma_i $~\citep{Fano:1983zz}, whose expectation values $\braket{\vec{\sigma}\cdot\vec{\sigma}}=-3$ and $1$ correspond to total spin quantum numbers of $0$ and $1$, respectively.
For hyperon pairs produced in electron-positron annihilation, it is straightforward to demonstrate that
\begin{equation}
\braket{\vec{\sigma}\cdot\vec{\sigma}}=\mathrm{Tr}(\rho_{Y\bar{Y}}\,\vec{\sigma}\cdot\vec{\sigma})=\mathrm{Tr}(\vec{C}_{L,T}) = 1.\label{eq:triplet}
\end{equation}
{regardless of the beam polarization degree ${P_{L,T}}$.}

\subsection{Quantum Discord and Steering}
\label{sec:quantum_discord}


Quantum discord is proposed as a measure of quantum correlations of a
bipartite quantum system by quantizing concepts from classical information theory \cite{Ollivier:2001fdq,Henderson:2001wrr}.
This paper focuses on one of categories based on measurement in any one of the subsystems which requires optimization over sets of
local measurements.

The quantum mutual information, describing the total correlation of a bipartite state $\rho_{AB}$ shared between two parties \textit{Alice} ($Y$) and \textit{Bob} ($\bar{Y}$), is defined as
\begin{equation} \label{eq:quantum}
  {I}(\rho_{AB}) = S(\rho_{A}) + S(\rho_{B}) - S(\rho_{AB}),
\end{equation}
where $S(\rho) = -\mathrm{Tr}(\rho \log_2 \rho)$ is the von Neumann entropy, and $\rho_{A(B)} = \mathrm{Tr}_{B(A)}(\rho_{AB})$ is the reduced density matrix of subsystem $A$ (or $B$) in Eq. \eqref{eq:marginal}.
Above definition is the quantum version of the classical mutual information obtained by replacing Shannon entropy by von Neumann entropy.

To evaluate the classical correlations of $\rho_{AB}$, ``classical'' mutual information is defined as
\begin{equation} \label{eq:classic}
{J}_{A}(\rho_{AB};\hat{n}) = S(\rho_A)-S_{A|B}^{\,\hat{n}},
\end{equation}
where $\hat{n}$ is the unit vector denoted the direction of local measurement for dichotomic observables ${\sigma}_{\hat{n}}=\hat{n}\cdot\vec{\sigma}$ safisfying ${\sigma}_{\hat{n}}|\pm\hat{n}\rangle=\pm|\pm\hat{n}\rangle$. The term $S_{A |B}^{\,\hat{n}}$ represents the expectation of the entropy for $\rho_{AB}$ following a local quantum measurement $\Pi_{\hat{n}}:=\mathbb{I}\otimes|\hat{n}\rangle\langle\hat{n}|$ on subsystem $B$:
\begin{equation}
S_{A|B}^{\,\hat{n}}=p_{\hat{n}} S(\rho_{\hat{n}}) + p_{-\hat{n}} S(\rho_{-\hat{n}}).
\end{equation}
Under a set of orthogonal projective measurement,
\begin{equation}
p_{\hat{n}} \rho_{\hat{n}} = \Pi_{\hat{n}} \rho_{AB} \Pi_{\hat{n}},\quad  p_{\hat{n}} = \mathrm{Tr}(\Pi_{\hat{n}} \rho_{AB} \Pi_{\hat{n}}).\label{eq:POVM}
\end{equation}
the classical mutual information is dependent on the choice of measurement basis.
Therefore, for a given bipartite state $\rho_{AB}$, the difference between two quantities given in Eq. \eqref{eq:quantum} and \eqref{eq:classic} was proposed to be a measure of quantum correlation, so called as quantum discord:
\begin{equation}
\mathscr{D}_{A}(\rho_{AB}) = {I}(\rho_{AB})-\max_{\hat{n}}{J_A}(\rho_{AB};\hat{n}), \label{eq:qd}
\end{equation}
satisfying $0 \leq \mathscr{D}_{A} \leq 1$.
Here the maximization is taken over all quantum measurements $\Pi_{\hat{n}}$ performed on the system $B$.
It can be expressed in a more simplified form as
\begin{equation}
  \begin{aligned}
  \mathscr{D}_{A}(\rho_{AB})&=S(\rho_{B})-S(\rho_{AB})+\min_{\hat{n}} \lf[p_{\hat{n}}S(\rho_{\hat{ n}})+p_{-\hat{n}}S(\rho_{-\hat{n}}) \rg]\\
  &=\;\min_{\hat{n}}S_{A|B}^{\,\hat{n}}-S_{A|B}, \label{eq:discord}
  \end{aligned}
\end{equation}
where quantum conditional entropy $S_{A|B}=S(\rho_{AB})-S(\rho_{B})$ is negative only when the state is entangled \cite{Kay:1998cm}.
Correspondingly, the $\min_{\hat{n}}S_{A|B}^{\,\hat{n}}$ is referred to {as the} classical conditional entropy.

{To calculate quantum conditional entropy $S_{A|B}$, it should be recalled that} the von Neumann entropy can be expressed as $S(\rho)=-\sum_{i}\lambda_{i}\log_2 \lambda_i$ with $\lambda_i$ being the eigenvalues of $\rho$.
Since the $\rho_{Y(\bar{Y})}^{P_{L,T}}$ are rank-$2$ matrices, the quantum conditional entropy can be formulated as:
\begin{equation}
  \begin{aligned}
    S(\rho_{Y}|\rho_{\bar{Y}})&=S(\rho_{Y\bar{Y}})-S(\rho_{\bar{Y}})
    =h(\lambda^{L,T})-h(\mu^{L,T}),\label{eq:qcs}
  \end{aligned}
\end{equation}
where $h(x)=-x\log_2 x - (1-x)\log_2 (1-x)$ is the Shannon binary entropy function, and $\lambda^{L,T}$, $\mu^{L,T}$ are eigenvalues given in Sec.\ref{sec:bfc} {for longitudinal and transverse beam, respectively}.

{The classical conditional entropy $\min_{\hat{n}}S_{A|B}^{\,\hat{n}}$ is calculated by explicitly writing} the local measurement in Eq.~\eqref{eq:POVM} as~\cite{Afik:2022dgh}:
\begin{equation}
 p_{\hat{n}}=\frac {1+\hat{n}\cdot{\vec{B}}^{-}}{2},\quad
\rho_{\hat{n}}=\frac{1}{2}\left[{1+\left(\frac{{\vec{B}}^{+}+{\vec{C}}\cdot\hat{n}}{1+\hat{ n}\cdot{\vec{B}}^{-}}\right)\cdot\boldsymbol{\sigma}}\right]\otimes{|\hat{n}\rangle\langle\hat{n}|}.\label{eq:condistate}
\end{equation}
{where the local measurement direction is parameterized without loss of generality as $\hat{n}=(\sin\vartheta_{\Pi}\cos\varphi_{\Pi},\sin\vartheta_{\Pi}\sin\varphi_{\Pi},\cos\vartheta_{\Pi})$.
$S_{A|B}^{\hat n}$ is then minimized over the solid angles $(\vartheta_{\Pi}, \varphi_{\Pi})$ using Eq.~\eqref{eq:condistate}. To ensure that we find the global minimum, we first perform a systematic scan over the measurement sphere and subsequently utilize the optimal grid points as seeds for a local minimization algorithm. This robust procedure is consistently applied to all numerical calculations of quantum discord presented in this work.
Analytical expressions for quantum discord are available and have been applied to top-antitop systems within the context of particle physics~\cite{Han:2024ugl}.
Those formulations are relatively straightforward as elementary particles are of vanishing polarization and
could be highly beneficial for event simulations by significantly accelerating the calculation of discord.
The analytical expressions for quantum discord remain
formidably complex given the specific density matrix considered here. We have therefore focused on numerical calculations in this study.}

As can be seen by Eq.~\eqref{eq:condistate} the local measurement on one subsystem $B$ ``steers'' the quantum state of other distant subsystem $A$.
Steering refers to the impossibility of describing the conditional states of one party using a local hidden state (LHS) model, suggested by Schr\"{o}dinger as a hypothesis for steering in place of local hidden variables \cite{Schrodinger:2008pyl}.
LHS model states that subsystem $B$ has a definite state, regardless of whether it is known to $B$, so that steering would remain experimentally undetectable \cite{Xiang:2022fng}.
The CJWR inequality \cite{Cavalcanti:2009wia}, a well-established criterion for testing nonlocal correlations, is herein leveraged to determine whether
hyperon pairs produced by polarized electron–positron annihilation exhibit quantum steering.
For a two-qubit system, the inequality with $3$-setting measurements on each subsystem is given by
\begin{equation}
F_{3} \equiv \frac{1}{\sqrt{3}} \left| \sum_{k=1}^{3} \mathrm{Tr}\left[ \rho_{AB} \left( A_{k} \otimes B_{k} \right) \right] \right| \leq 1,
\label{eq:CJWR}
\end{equation}
where $A_{k}=\vec{u}_{k}\cdot\boldsymbol{\sigma}$, $B_{k}=\vec{v}_{k} \cdot \boldsymbol{\sigma}$,
$\vec{u}_{k}$ and $\vec{v}_{k}$ (with $k = 1, 2, 3$) are unit vectors,
and $\{ \vec{v}_{1}, \vec{v}_{2}, \vec{v}_{3} \}$ form a Cartesian coordinate system within which any two vectors are mutually orthogonal.
By choosing the unit vectors $\vec{v}_{k}$ such that
\be
\| \vec{C} \vec{v}_{k} \|  = \sqrt{ \mathrm{Tr}(\vec{v}_{k}^\mathrm{T} \vec{C}^\mathrm{T}\vec{C}\vec{v}_{k}) } = \sqrt{ \frac{1}{3}{\mathrm{Tr}(\vec{C}^\mathrm{T}\vec{C})}},
\ee
and setting $\vec{u}_{k} = \vec{C} \vec{v}_{k}\,/\, \| \vec{C} \vec{v}_{k} \|$,
the quantity $F_{3}$ in Eq.~\eqref{eq:CJWR} attains its maximum, labeled as the CJWR parameter $\mathcal{F}_3$~\cite{Costa:2016pas}:
\begin{equation}
\mathcal{F}_{3}[\rho_{AB}] := \max_{\vec{u}_{k},\vec{v}_{k}} F_{3}= \|\vec{C}\|=\sqrt{ \mathrm{Tr}\left( \vec{C}^{\mathrm{T}} \vec{C} \right) },\label{eq:F3}
\end{equation}
where $\vec{C}$ is the correlation matrix in Eq. \eqref{eq:CL} or \eqref{eq:CT} for hyperon pair systems.
Similar to the violations of the Clauser-Horne-Shimony-Holt (CHSH) inequality, a violation of CJWR steering inequality implies a nonlocal property of the quantum system, thus  ruling out any LHS model \cite{Cavalcanti:2009wia,Costa:2016pas}.
When $\mathcal{F}_{3}[\rho_{AB}]$ lies within $(1, \sqrt{3}]$, the bipartite system $\rho_{AB}$ is refered as CJWR-steerable and $\mathcal{F}_{3} = \sqrt{3}$ is maximal CJWR-steerable.

Quantum discord and steering are both measures of quantum correlations defined through local measurements, capturing distinct aspects of non-classicality from informational and operational perspectives, respectively.
Unlike entanglement of formation and Bell nonlocality, their definitions imply an inherent asymmetry in quantum correlations of bipartite system.
However, since $CP$ violation is not considered in Sec.~\ref{sec:bfc}, discord and steering are symmetric with respect to the two subsystems.

\section{Longitudinal Beam Polarization}
\label{sec:PL}

\begin{figure}[!ht]
\centering
	\includegraphics[width = 0.75 \linewidth]{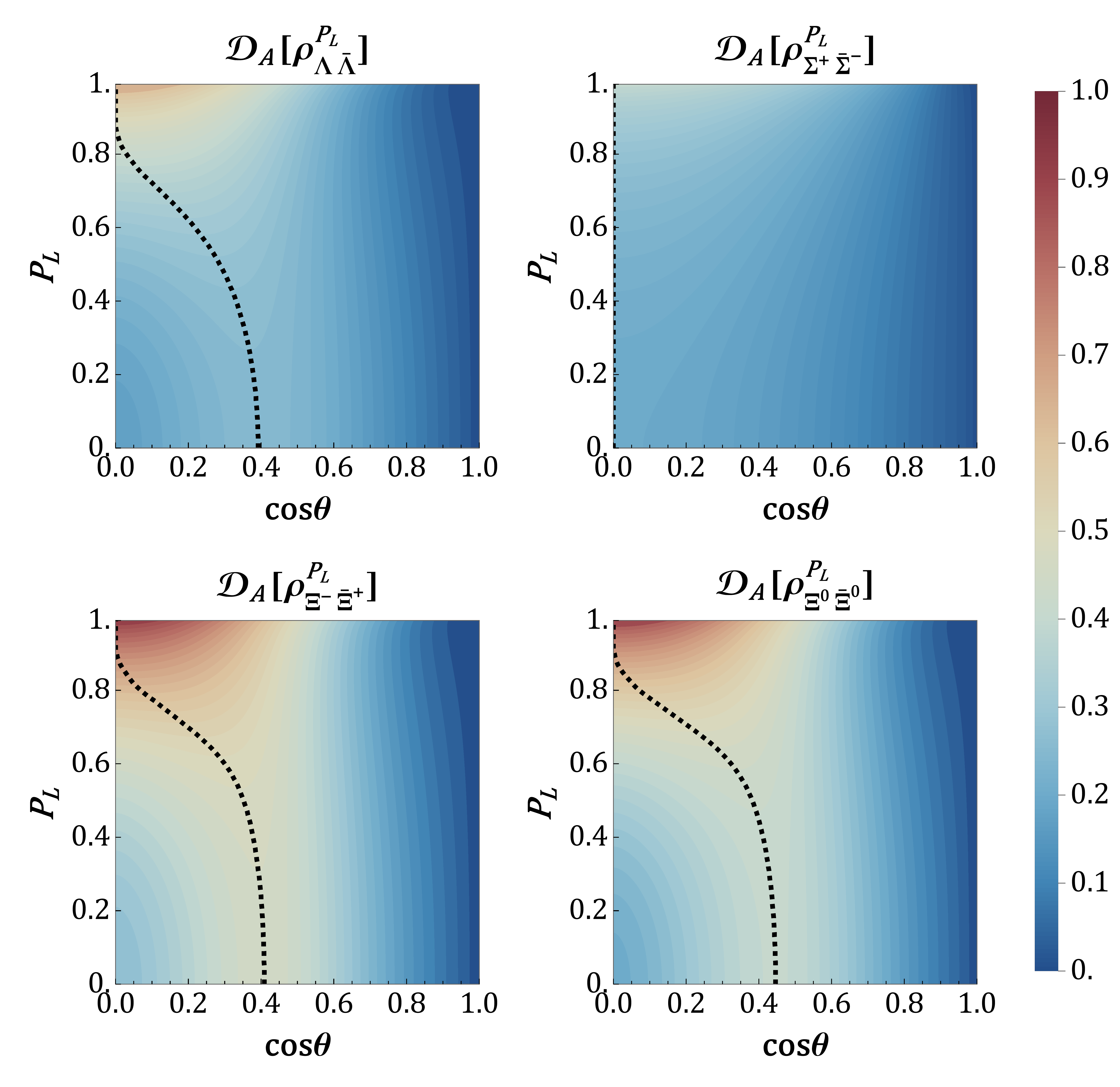}
\caption{\raggedright
The quantum discord $\mathscr{D}_{A}[\rho^{P_L}_{Y\bar{Y}}]$ as a function of $\cos\theta$ and longitudinal polarization degree $P_L$ in $J/\psi\to Y{\bar{Y}}$ for $Y=\Lambda$, $\Sigma^+$, $\Xi^{-}$ and $\Xi^{0}$ respectively.
See main text for the explanation of the dashed curves.
} \label{fig:DPL1}
\end{figure}

\begin{figure}[!htb]
\centering
  \includegraphics[width = 0.65 \linewidth]{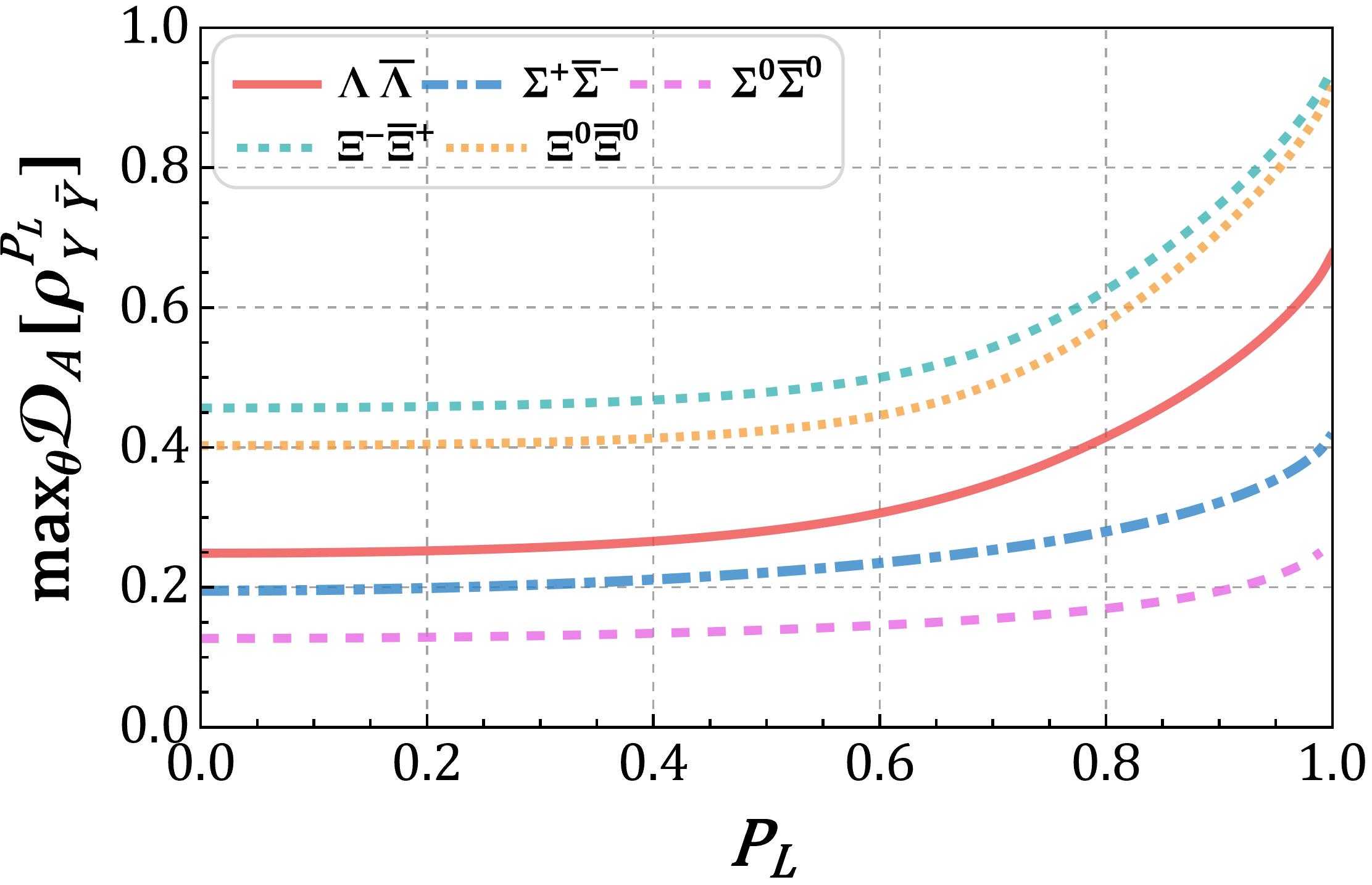}
\caption{\raggedright
The $\max_{\theta}\mathscr{D}_{A}[\rho_{Y\bar{Y}}^{P_L}]$ as a function of $P_L$ in $J/\psi\to Y{\bar{Y}}$ for $Y=\Lambda$, $\Sigma^+$, $\Sigma^0$, $\Xi^{-}$ and $\Xi^{0}$ respectively.
} \label{fig:DPL2}
\end{figure}

The effect of longitudinal polarization of lepton beam on quantum discord of hyperon pairs is illustrated in Fig.~\ref{fig:DPL1} based on the parameters in Table~\ref{tab:decay_parameters}.
Here and following the results of $J/\psi \rightarrow Y\bar{Y}$ decay for $Y = \Lambda$, $\Sigma^{+}$, $\Xi^{-}$, and $\Xi^{0}$ are shown in main text, see Appendix.~\ref{sec:Sigma0} for $Y=\Sigma^{0}$.
Note that the discord $\mathscr{D}_{A}[\rho^{P_L}_{Y\bar{Y}}]$ exhibits the symmetries $\mathscr{D}_{A}(\theta) = \mathscr{D}_{A}(\pi - \theta)$ with $\theta \in [0,\pi]$, $\mathscr{D}_{A}(P_L) = \mathscr{D}_{A}(-P_L)$ with $P_L \in [-1,1]$. The CJWR parameter $\mathcal{F}_3[\rho^{P_L}_{Y\bar{Y}}]$ shares these symmetries. Consequently, we only plot the $\cos\theta$ and $P_L$ dependence for $P_L \in [0,1]$ and $\theta \in [0,\pi/2]$.

For $\Lambda\bar{\Lambda}$, $\Xi^{-}\bar{\Xi}^{+}$ and $\Xi^{0}\bar{\Xi}^{0}$,
the dashed curves in Fig.~\ref{fig:DPL1} represent a specific scattering angle, labeled as ${\theta}_{\max}$, corresponding to the largest value of $\mathscr{D}_{A}$ at fixed $P_L$, denoted as $\max_{\theta}\mathscr{D}_A$.
It can be seen that the $\theta_{\max}$ shifts toward the angle perpendicular to the beam direction as $P_L$ increases, reaching $\theta = \pi/2$ at the maximum degree of beam polarization.
Within interval of $(\theta_{\max},\pi-\theta_{\max})$, $\mathscr{D}_{A}$ grows monotonically with \(P_L\), exhibiting relatively high sensitivity to beam polarization degree.
For $\Sigma^{+}\bar{\Sigma}^{-}$, $\Sigma^{0}\bar{\Sigma}^{0}$ pairs, increasing $P_L$ enhances $\mathscr{D}_{A}$ over the full range of $\theta$, except at $\theta=0$ and $\pi$.
The maximum value, $\max_{\theta}\mathscr{D}_{A}$, consistently occurs at $\theta = \pi/2$ for all $P_L\in[0,1]$.

For all channels as shown in Fig.~\ref{fig:DPL2}, $\max_{\theta}\mathscr{D}_{A}$ increases with larger \(P_L\), but stays below 1, falling short of the theoretical left limit.
The $\max_{\theta}\mathscr{D}_{A}$ of $\Sigma^{+}\bar{\Sigma}^{-}$ and $\Sigma^{0}\bar{\Sigma}^{0}$ pairs are insensitive to $P_L$, as illustrated by the following analytical formula.
At $\theta = 0,\pi/2,\pi$, the classical conditional entropy in Eq.~\eqref{eq:discord} vanishes, e.g. $\min_{\hat{n}} S_{A | B}^{\,\hat{n}}\equiv 0$,
leading to $\mathscr{D}_{A}(\theta=0,\frac{\pi}{2},\pi)=-S_{A | B}(\theta=0,\frac{\pi}{2},\pi)$.
From Eqs.~\eqref{eq:eigenPL}, \eqref{eq:rdmeigenPL} and \eqref{eq:qcs}, we have
\begin{equation}
  \begin{aligned}
 \mathscr{D}_{A}(\theta &= \frac{\pi}{2}) = h\left[\frac{1}{2} \left(1+\gamma _{\psi} P_L\right)\right]\\
 &\qquad\;\;\,-h\left[\frac{1}{2} \left(1+\sqrt{\alpha _{\psi }^2+(1-\alpha _{\psi }^2) P_L^2}\right)\right];\\
 \mathscr{D}_{A}(\theta &= 0,\pi) \equiv 0.
  \end{aligned} \label{eq:disangle}
\end{equation}
holding for all channels.
For $\Sigma^{+}\bar{\Sigma}^{-}$ and $\Sigma^{0}\bar{\Sigma}^{0}$, $\max_{\theta}\mathscr{D}_{A} = \mathscr{D}_{A}(\theta = \frac{\pi}{2})$ exhibits little dependence on $P_L$ due to the cancellation in Eq. \eqref{eq:disangle}, which stems from the small relative phase between electric and magnetic couplings in Table \ref{tab:decay_parameters}, yielding $\gamma_{\psi} \simeq \sqrt{1-\alpha_{\psi }^2}$.
Likewise, the same small relative phase also leads to the insensitivity of $\max_{\theta}\mathcal{B}$, $\max_{\theta}\mathcal{C}$ and $\max_{\theta}\mathcal{N}$  (for the CHSH parameter, concurrence, and negativity, respectively) to $P_L$, as their dependence involves the term $\beta_{\psi} P_L$  (with $\beta_{\psi} \simeq 0$) for $\Sigma^{+}\bar{\Sigma}^{-}$ and $\Sigma^{0}\bar{\Sigma}^{0}$ \cite{Zhang:2026nwm}.

\begin{figure}[!ht]
\centering
  		\includegraphics[width = 0.75\linewidth]{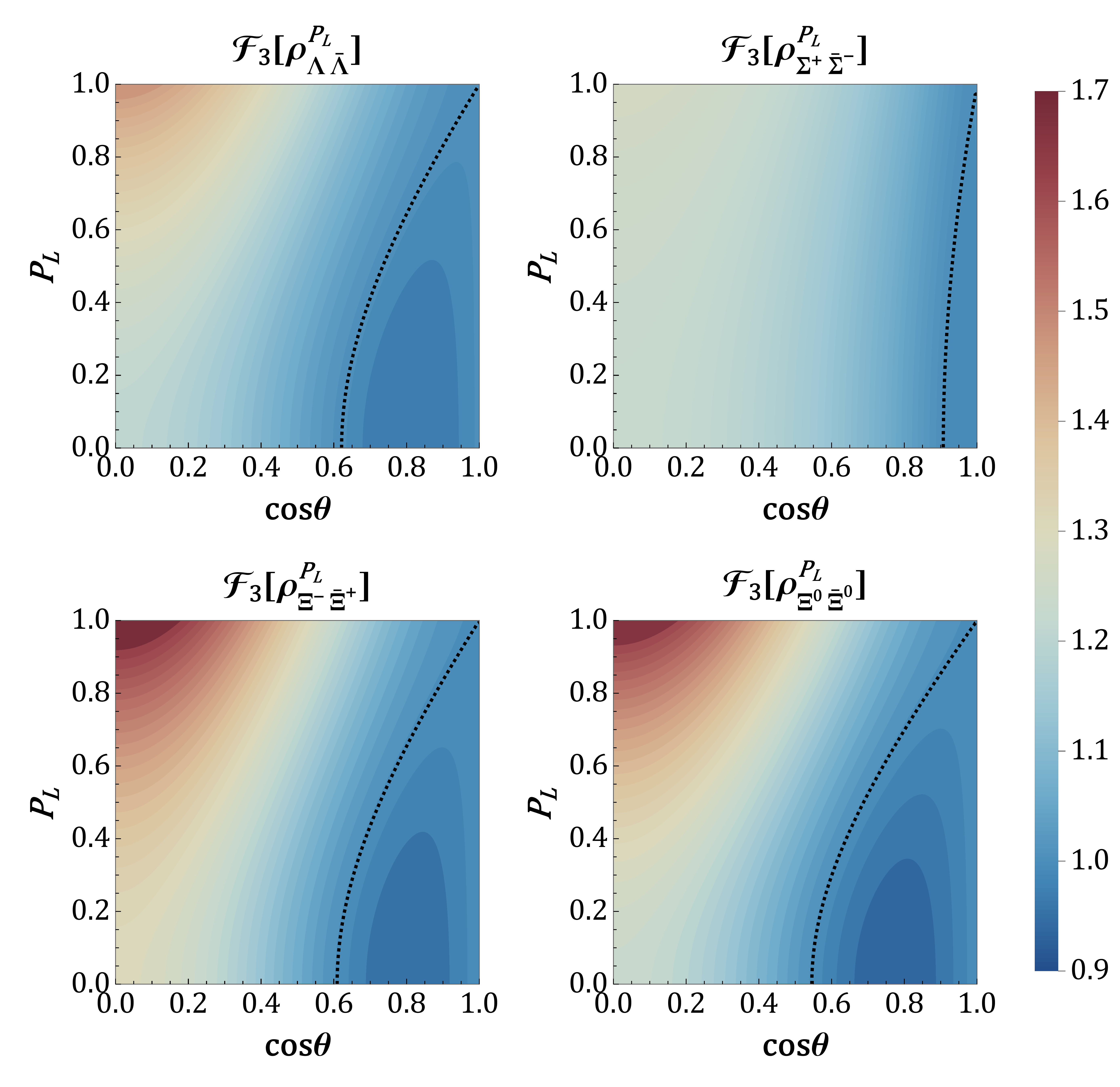}
\caption{\raggedright
The CJWR parameter $\mathcal{F}_{3}[\rho_{Y\bar{Y}}^{P_L}]$ as a function of $\cos\theta$ and $P_L$ in $J/\psi\to Y{\bar{Y}}$ for $Y=\Lambda$, $\Sigma^+$, $\Xi^{-}$ and $\Xi^{0}$ respectively. The black dashed curve is for $\mathcal{F}_{3}[\rho_{Y\bar{Y}}^{P_L}] = 1$.
} \label{fig:F_PL}
\end{figure}
\begin{figure}[!ht]
\centering
  		\includegraphics[width = 0.49 \linewidth]{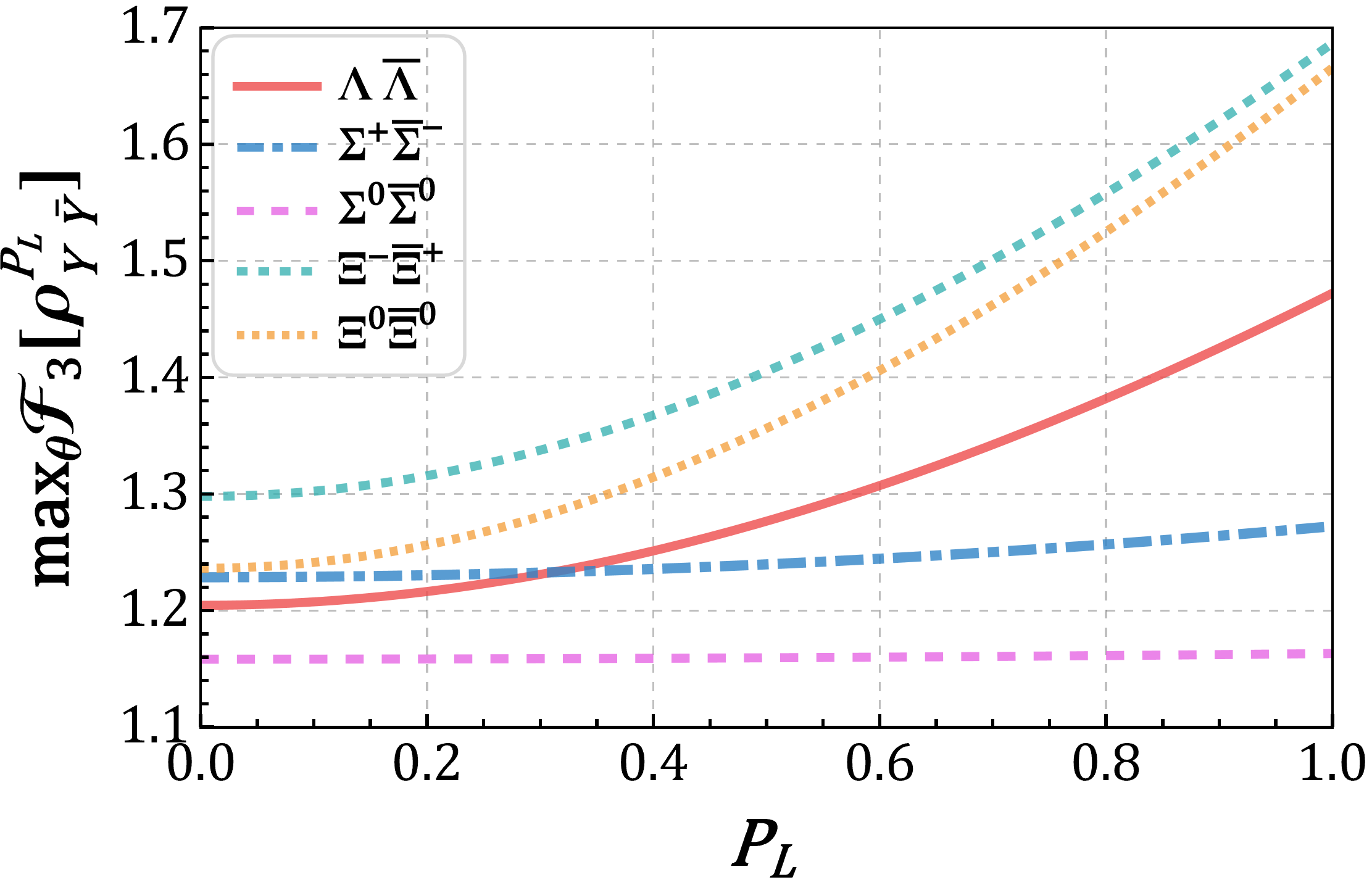}
      \includegraphics[width = 0.49 \linewidth]{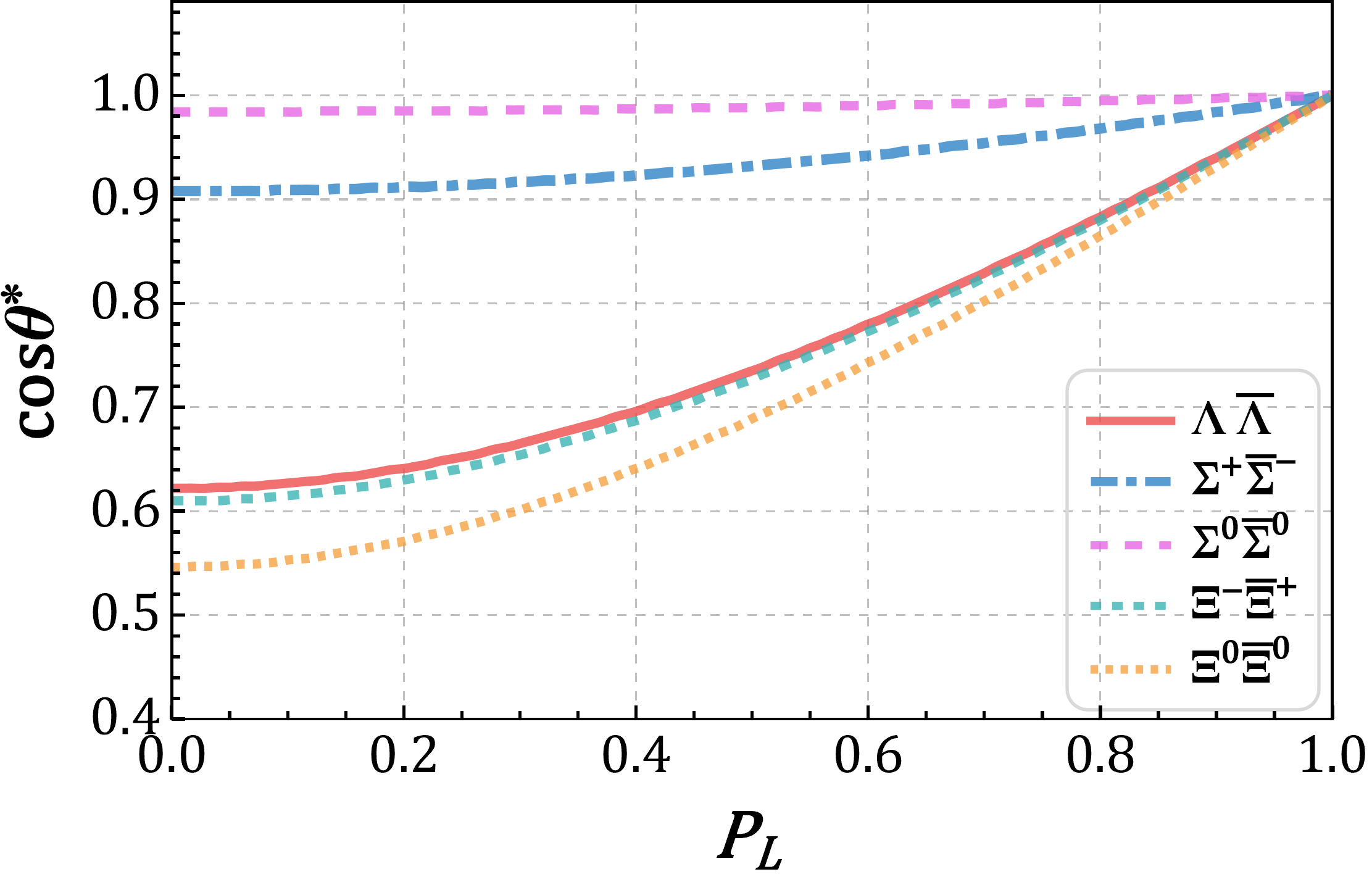}
\caption{\raggedright
The $\max_{\theta}\mathcal{F}_{3}[\rho_{Y\bar{Y}}^{P_L}]$ (left panel) and the $\cos\theta^{*}$ (right panel) as a function of $P_L$ in $J/\psi\to Y{\bar{Y}}$ for $Y=\Lambda$, $\Sigma^+$, $\Sigma^0$, $\Xi^{-}$ and $\Xi^{0}$ respectively.
} \label{fig:maxF_PL}
\end{figure}

Fig.~\ref{fig:F_PL} displays the $\cos\theta$ and $P_L$ dependence of the CJWR parameter $\mathcal{F}_{3}[\rho^{P_L}_{Y \bar Y}]$. It can be seen that increasing $P_L$ enhances $\mathcal{F}_{3}$ across the full range of $\theta$ with an exception of $\theta=0$ and $\pi$, under which angles $\mathcal{F}_{3}(\theta = 0,\pi) = 1$, see Eqs.~\eqref{eq:CL} and \eqref{eq:F3}.
In Fig.~\ref{fig:F_PL}, the left areas of dashed curve in each panel represents CJWR steerable $\mathcal{F}_3\in(1, \max_{\theta}\mathcal{F}_3]$ with $\max_{\theta}\mathcal{F}_3$ being the largest CJWR parameter with respect to $\theta$.
From Eqs.~\eqref{eq:CL} and \eqref{eq:F3}, the $\max_{\theta}\mathcal{F}_{3}$ locates at $\theta = \pi/2$ with any $P_L$,
\begin{align}
\max_{\theta}\mathcal{F}_{3}=\mathcal{F}_{3}(\theta = \frac{\pi}{2}) = \sqrt{1 + 2\alpha_\psi^2 + 2{P_L^2}\beta_\psi^2 }<\sqrt{3}, \label{eq:F3PL1}
\end{align}
which is showed in the left panel of Fig.~\ref{fig:maxF_PL}.

Increasing $P_L$ expands the angular coverage of steerability as can be also seen in Fig.~\ref{fig:F_PL}.
The corresponding angle for $\mathcal{F}_3 = 1$ is defined as the critical angle $\theta^*$, shown in the right panel of Fig.~\ref{fig:maxF_PL}.
The $|\cos\theta^*|$ moves to 1 when $P_L$ is approaching the maximum degree 1, so that hyperon pairs is always CJWR-steerable across the scattering angular range $(0,\pi)$ at $P_L = 1$.
It is noted that for $\Sigma^{+}\bar{\Sigma}^{-}$ and $\Sigma^{0}\bar{\Sigma}^{0}$ pairs $\mathcal{F}_3$ exhibits relatively low dependence on $P_L$, and $\theta^*$ is already close to zero in the case of unpolarized beams.

\section{Transverse Beam Polarization}
\label{sec:PT}
\begin{figure}[!t]
\centering
  		\includegraphics[width = 0.75\linewidth]{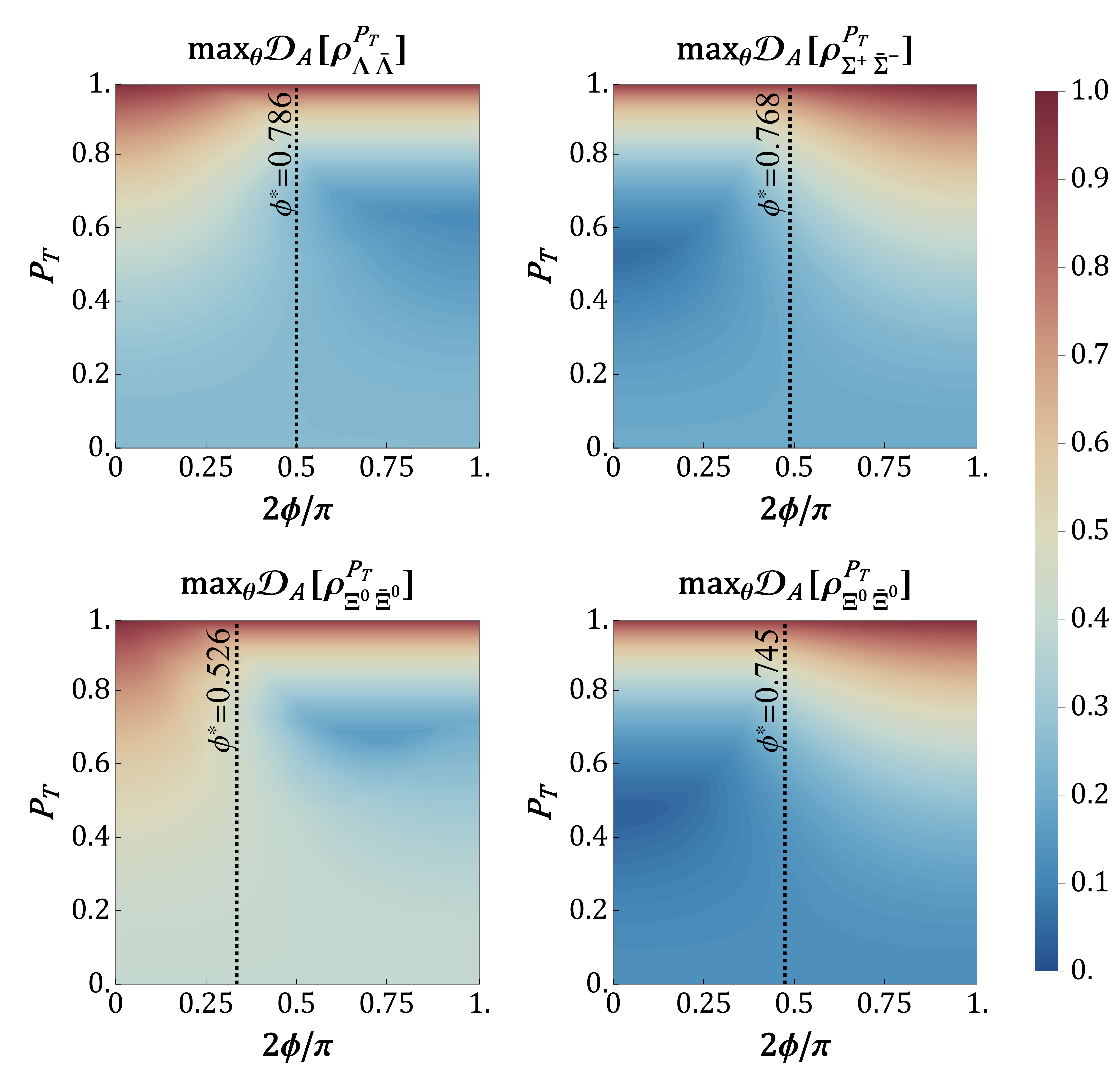}
\caption{\raggedright
The $\max_{\theta}\mathscr{D}_{A}[\rho_{Y\bar{Y}}^{P_T}]$ as a function of azimuthal angle $\phi$ and transverse polarization of beams $P_T$ in $J/\psi\to Y{\bar{Y}}$ for $Y=\Lambda$, $\Sigma^+$, $\Xi^{-}$ and $\Xi^{0}$ respectively. The black dashed line is for critical azimuthal angle $\phi^*$ in rad, see main text for more details.} \label{fig:DPT1}
\end{figure}

\begin{figure}[!t]
\centering
        \includegraphics[width = 0.65 \linewidth]{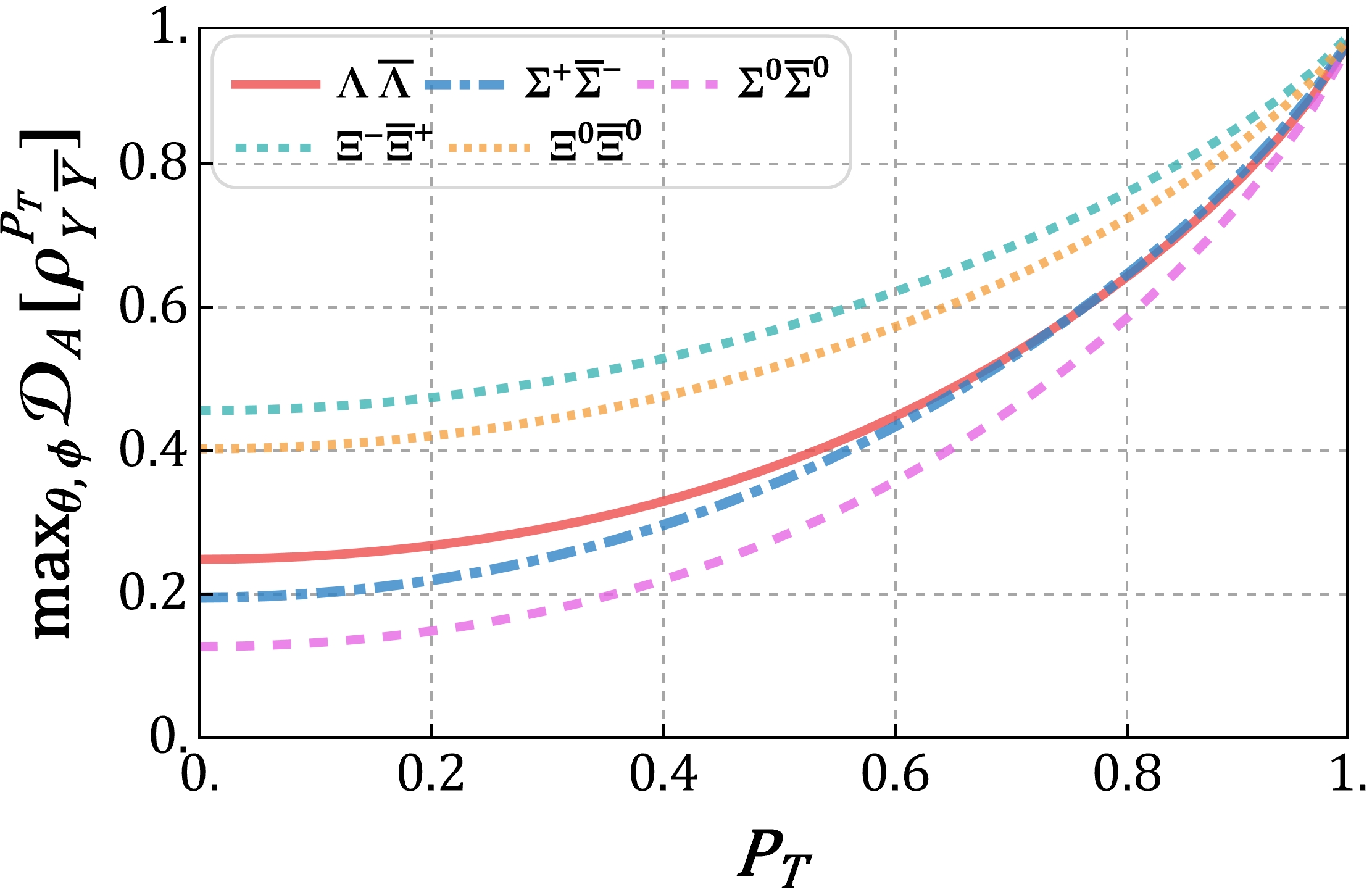}
\caption{\raggedright
The $\max_{\theta,\phi}\mathscr{D}_{A}[\rho_{Y\bar{Y}}^{P_T}]$ as a function of $P_T$ in $J/\psi\to Y{\bar{Y}}$ for $Y=\Lambda$, $\Sigma^+$, $\Sigma^0$, $\Xi^{-}$ and $\Xi^{0}$ respectively.
} \label{fig:DPT2}
\end{figure}

The case of transverse polarized beam is more complicated considering that it introduces the dependence on the azimuthal angle $\phi$ perpendicular to the scattering plane.
We will search for the largest discord and CJWR parameter along the  $\theta$-axis at a fixed $\phi$ angle.
Note that discord maintains the symmetries $\mathscr{D}_{A}(\theta) = \mathscr{D}_{A}(\pi - \theta)$ with $\theta \in [0,\pi]$,  $\mathscr{D}_{A}(P_T) = \mathscr{D}_{A}(-P_T)$ with $P_T \in [-1,1]$, $\mathscr{D}_{A}(\phi)=\mathscr{D}_{A}(\pi-\phi)$, $\mathscr{D}_{A}(\phi)=\mathscr{D}_{A}(\phi+\pi)$ with $\phi \in [0,\pi]$. The CJWR parameter $\mathcal{F}_3[\rho^{P_T}_{Y\bar{Y}}]$ holds the same symmetries.

The $P_T$ and $\phi$ dependence of the largest $\mathscr{D}_A[\rho_{Y\bar{Y}}^{P_T}]$ with respect to $\theta$, labeled as $\max_{\theta}\mathscr{D}_A[\rho_{Y\bar{Y}}^{P_T}]$ are showed in Fig.~\ref{fig:DPT1}.
The dashed line is a critical $\phi^*$ dividing each panel into two regions.
For $\Lambda\bar{\Lambda}$, $\Xi^{-}\bar{\Xi^{-}}$, $\Xi^{0}\bar{\Xi^{0}}$, the $\max_{\theta}\mathscr{D}_{A}$ monotonously increases as the increasing of $P_T$ in the left region of the dashed line, e.g. $\phi\in[0,\phi^*]$.
For $\Sigma^{+}\bar{\Sigma^{+}}$ and $\Sigma^{0}\bar{\Sigma^{0}}$, the $\max_{\theta}\mathscr{D}_{A}$ monotonously increases
as the increasing of $P_T$ in the right side of the dashed line, e.g. $\phi\in[\phi^*,\pi/2]$.

When $\theta = 0,\pi/2,\pi$, the classical conditional entropy in Eq.~\eqref{eq:discord} $\min_{\hat{n}} S_{A | B}^{\,\hat{n}}[\rho_{Y\bar{Y}}^{P_T}] = 0$, so that $\mathscr{D}_{A}(\theta=0,\frac{\pi}{2},\pi)=-S_{A | B}(\theta=0,\frac{\pi}{2},\pi)$. From Eq.~(\ref{eq:qcs}), we have
\begin{align}
    \nonumber
 \mathscr{D}_{A}(\theta = \frac{\pi}{2}) = & h\left[\frac{1}{2}+\frac{P_T^2\beta_{\psi} \sin 2\phi}{2+ 2 P_T^2 \alpha _{\psi }  \cos 2 \phi }\right]\\
 -&h\left[\frac{1}{2}+\frac{\sqrt{\alpha _{\psi }^2+2 P_T^2 \alpha _{\psi } \cos 2 \phi + P_T^4 \left(1-\alpha _{\psi }^2 \sin ^2 2 \phi \right)}}{2+ 2 P_T^2 \alpha _{\psi }  \cos 2 \phi }\right],\\
 \mathscr{D}_{A}(\theta = 0,\pi) = & 1-h\left[\frac{1}{2} \left(1+P_T^2\right)\right]. \label{eq:DPT}
\end{align}
Note that $\mathscr{D}_{A}(\theta = 0,\pi) $ in Eq.~\eqref{eq:DPT} is independent of $\phi$.
Unlike longitudinal beam polarization, there is no definite functional relationship between $\mathscr{D}_{A}(\theta=0,\frac{\pi}{2},\pi)$ and $\max_{\theta,\phi}\mathscr{D}_{A}$.

As can be seen in Fig.~\ref{fig:DPT1}, the largest $\max_{\theta}\mathscr{D}_{A}$, labeled as $\max_{\theta,\phi}\mathscr{D}_{A}$, is attained at specific azimuthal angle $\phi$ when $P_T$ holds constant.
As shown in Fig.~\ref{fig:DPT2}, the $\max_{\theta,\phi}\mathscr{D}_{A}$ is always enhanced as the increasing of $P_T$, and occurs at specific angles of
\begin{itemize}
  \item[a.] $\phi=k\pi \ (k=0,1,2)$ for $\Lambda\bar{\Lambda}$, $\Xi^{-}\bar{\Xi^{-}}$, $\Xi^{0}\bar{\Xi^{0}}$ pairs;
  \item[b.] $\phi=\pi/2+ k\pi \ (k=0,1)$ for $\Sigma^{+}\bar{\Sigma^{+}}$, $\Sigma^{0}\bar{\Sigma^{0}}$ pairs.
\end{itemize}

\begin{figure}[!t]
\centering
  		\includegraphics[width = 0.75\linewidth]{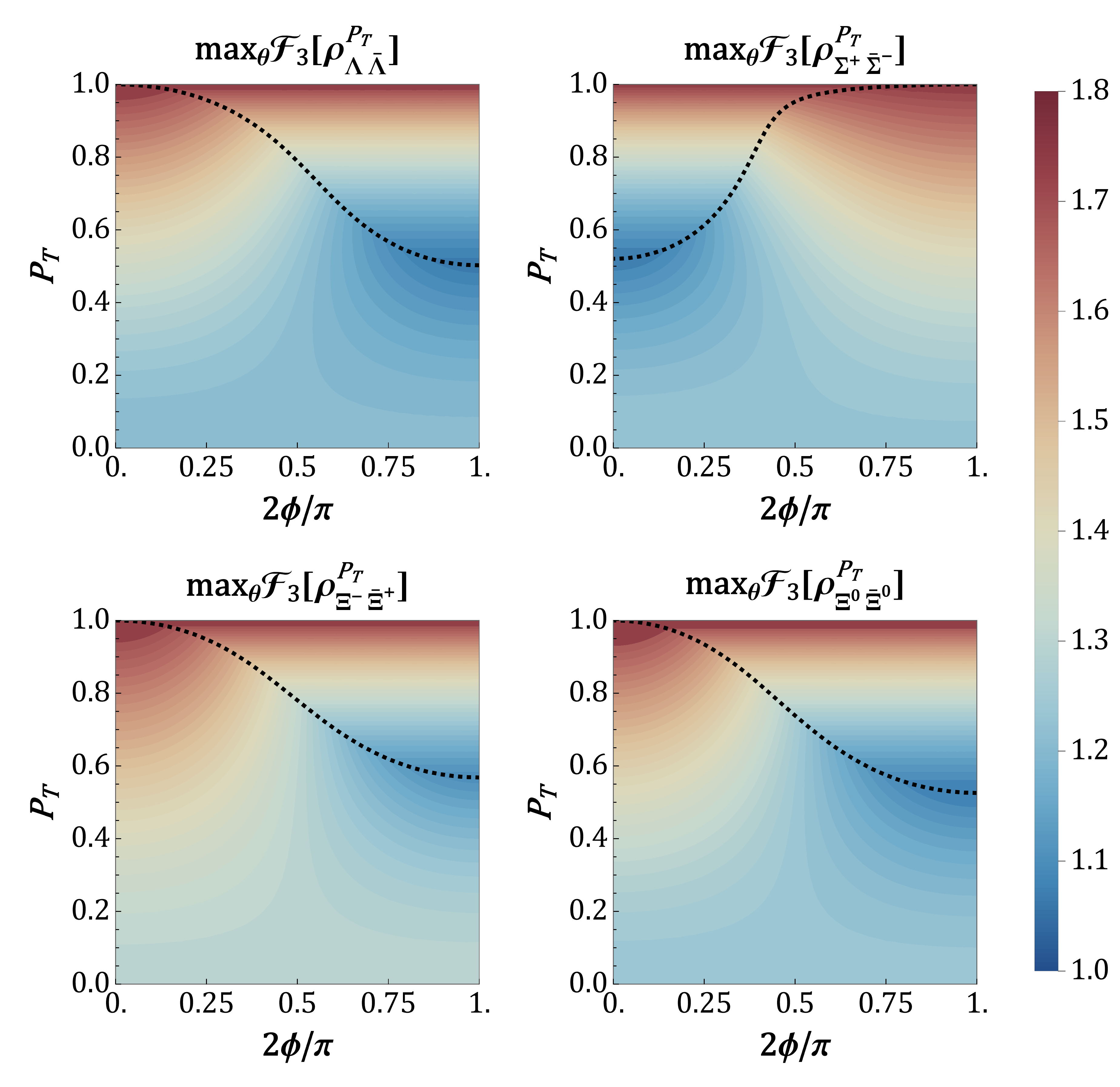}
\caption{\raggedright
The $\max_{\theta}\mathcal{F}_{3}[\rho_{Y\bar{Y}}^{P_T}]$ as a function of $\phi$ and $P_T$ in $J/\psi\to Y{\bar{Y}}$ for $Y=\Lambda$, $\Sigma^+$, $\Xi^{-}$ and $\Xi^{0}$ respectively. The black dashed curve is $\mathcal{F}_{3}(\theta=0,\pi)=\mathcal{F}_{3}(\theta=\pi/2)$. For further details of the curve, refer to the main text.
} \label{fig:FPT1}
\end{figure}

\begin{figure}[!t]
\centering
      \includegraphics[width = 0.51 \linewidth]{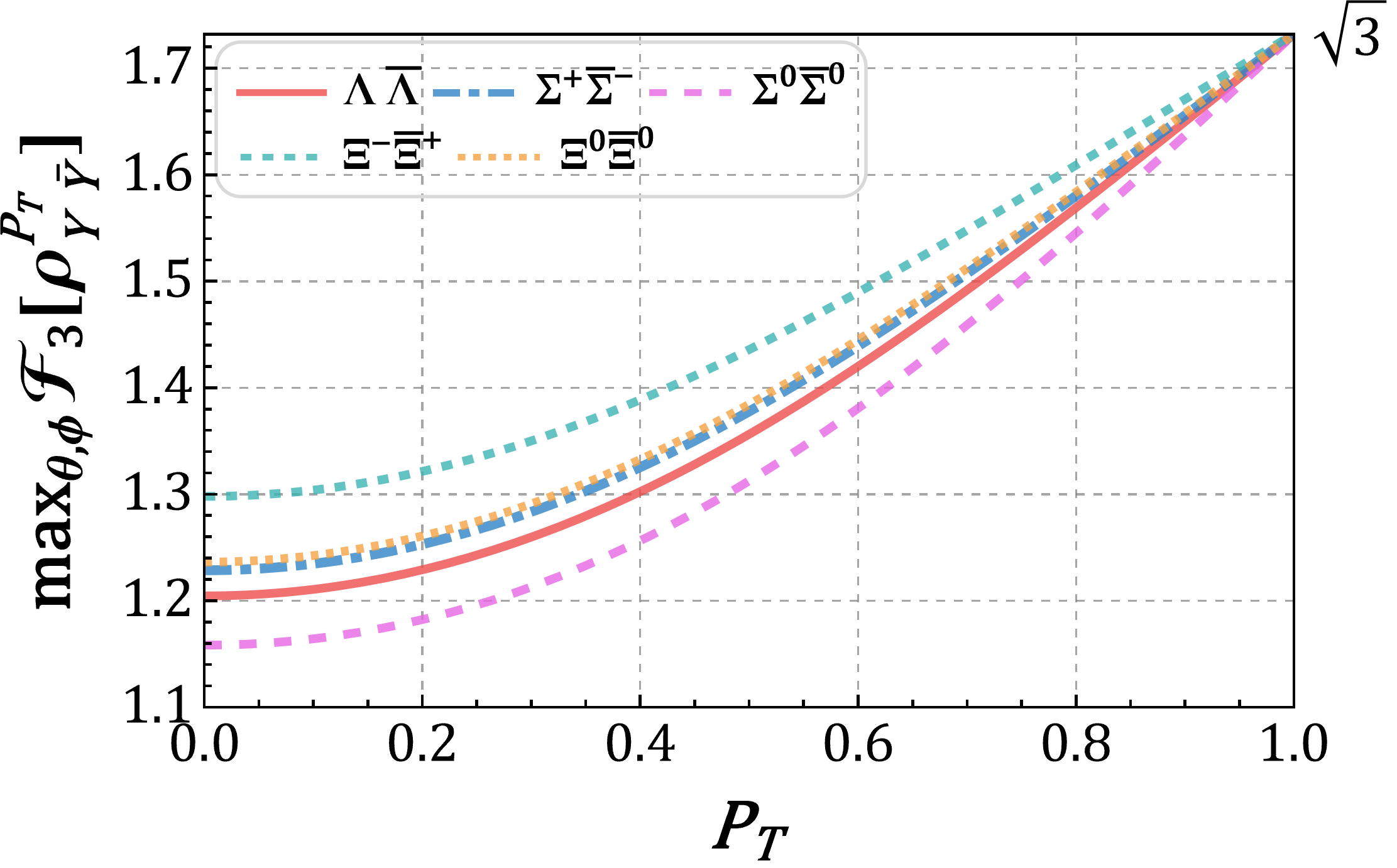}
  	\includegraphics[width = 0.48 \linewidth]{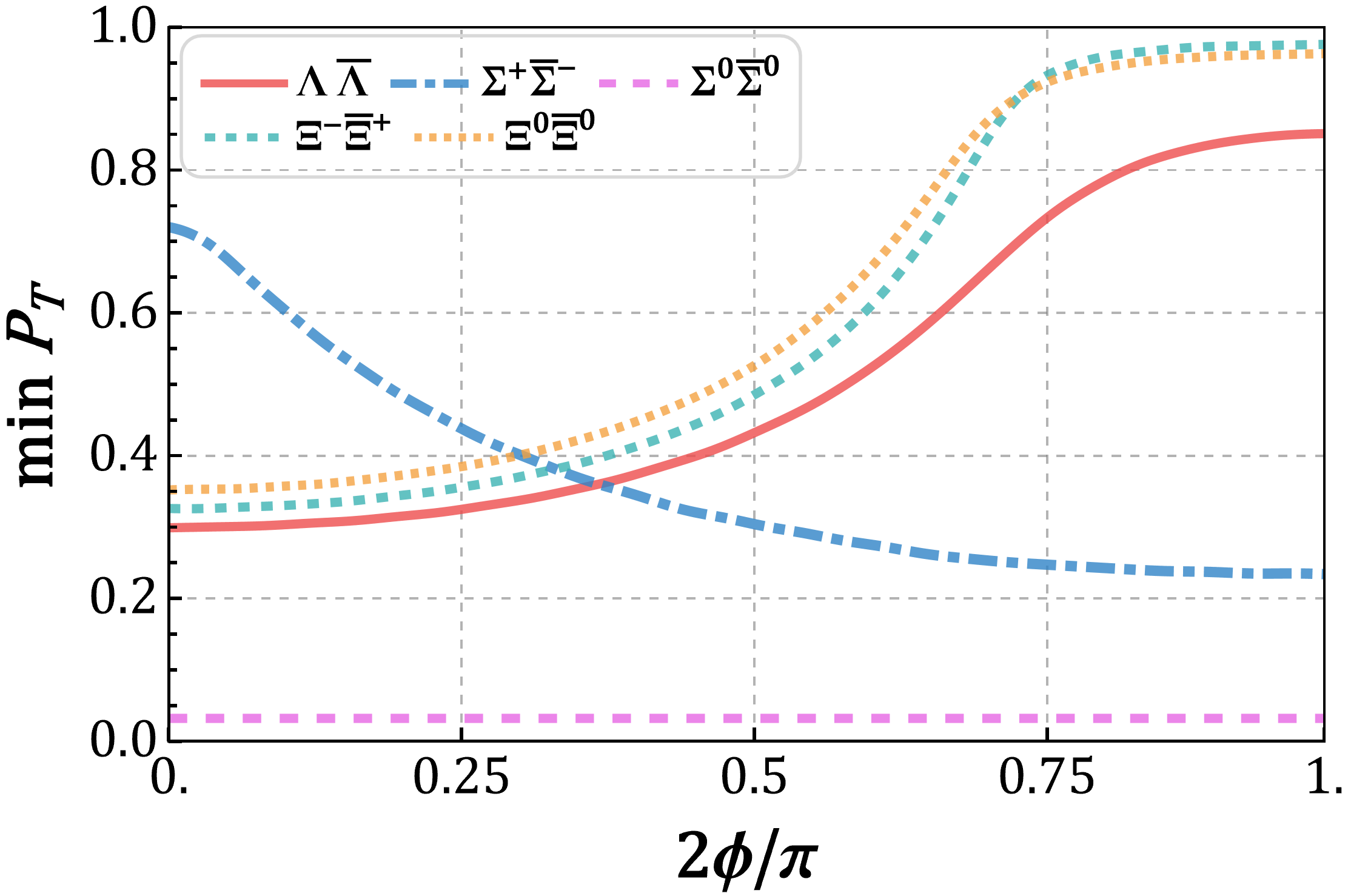}
\caption{\raggedright
The $\max_{\theta,\phi}\mathcal{F}_{3}[\rho_{Y\bar{Y}}^{P_T}]$ (left) as a function of $P_T$, and the $\min P_T$ (right) as a function of azimuthal angle $\phi \in [0,\pi/2]$ in $J/\psi\to Y{\bar{Y}}$ for $Y=\Lambda$, $\Sigma^+$, $\Sigma^0$, $\Xi^{-}$ and $\Xi^{0}$ respectively. See main text for the definition of $\min P_T$.
} \label{fig:FPT2}
\end{figure}

Fig.~\ref{fig:FPT1} displays the $P_T$ and $\phi$ dependence of the largest CJWR parameter with respect to $\theta$, labeled as $\max_{\theta}\mathcal{F}_{3}[\rho^{P_L}_{Y \bar Y}]$.
The $\max_{\theta}\mathcal{F}_3$ happens either at scattering angle of $\theta=0,\pi$ or $\theta={\pi}/{2}$ when $P_T\in[0,1)$, which can be derived from Eq.~\eqref{eq:CT} and \eqref{eq:F3}:
\begin{align}
\mathcal{F}_3(\theta =& 0,\pi)  =  \sqrt{1+2P_T^4},\label{eq:F3PT1} \\
\mathcal{F}_3(\theta =& \frac{\pi}{2})  =  \frac{\sqrt{1+2 \alpha _{\psi }^2 +6  P_T^2\alpha _{\psi } \cos 2 \phi +P_T^4\varepsilon}}{1+ P_T^2 \alpha_\psi \cos 2\phi }, \label{eq:F3PT2}
\end{align}
with $\varepsilon=2 \cos ^2 2 \phi +\alpha_{\psi }^2 \cos ^2 2 \phi +2 \gamma _{\psi }^2 \sin ^2 2 \phi $.
When $P_T=1$, the $\max_{\theta}\mathcal{F}_3$ occurs at
\begin{itemize}
  \item[a.] $\theta=0,\pi$ and $\theta=\pi/2$ simultaneously if $\phi=\pi/2+k\pi$ $(k=0,1)$;
  \item[b.] full $\theta$ range if $\phi=k\pi$ $(k=0,1,2)$;
  \item[c.]  otherwise, $\theta=0, \pi$ or $\theta= \pi/2$ as determined by Eqs.~(\ref{eq:F3PT1},\ref{eq:F3PT2}).
\end{itemize}

The dashed curve $\mathcal{F}_{3}(\theta=0,\pi)=\mathcal{F}_{3}(\theta=\pi/2)$ in Fig.~\ref{fig:FPT1} divides each panel into two regions.
For the region above the dashed curve, the $\max_{\theta}\mathcal{F}_3=\mathcal{F}_3(\theta = 0,\pi)$, and attain maximal violation of CJWR inequality for arbitary $\phi$ at $P_T=1$ (see Eq.~\eqref{eq:F3PT1}).
For the region below the dashed curve, the $\max_{\theta}\mathcal{F}_3=\mathcal{F}_3(\theta = \pi/2)$, and hold the maximal violation for $\phi=k\pi/2 \ (k=0,1,2,3,4)$ at $P_T=1$ (see Eq.~\eqref{eq:F3PT2}).
Futhermore, the maximal violation is attained across the full $\theta$ range for $\phi=k\pi \ (k=0,1,2)$ at $P_T=1$.
It is found for both regions $\max_{\theta}\mathcal{F}_3 \geq 1$ across full $\phi$ range, and $P_T = 1$ drives $\max_{\theta}\mathcal{F}_3[\rho^{P_T}_{Y\bar{Y}}]$ approach $\sqrt{3}$ at any fixed $\phi$.

As can be seen in Fig.~\ref{fig:FPT1}, the largest $\mathcal{F}_3$ with respect to helicity angles $\{\theta,\phi\}$, denoted as $\max_{\theta,\phi}\mathcal{F}_3$, is attained as specific $\phi$ when $P_T$ holds constant.
As shown in left panel of Fig.~\ref{fig:FPT2}, the $\max_{\theta,\phi}\mathcal{F}_3$ is always enhanced as the increasing of $P_T$, and occurs at specific angles of
\begin{itemize}
  \item[a.] $\phi=k\pi \ (k=0,1,2)$ for $\Lambda\bar{\Lambda}$, $\Xi^{-}\bar{\Xi^{-}}$ and $\Xi^{0}\bar{\Xi^{0}}$ pairs;
  \item[b.] $\phi=\frac{\pi}{2}=k\pi \ (k=0,1)$ for $\Sigma^{+}\bar{\Sigma^{+}}$, $\Sigma^{0}\bar{\Sigma^{0}}$ pairs.
\end{itemize}

The violation of CJWR inequality across full $\theta$ range can be achieved by increasing the transverse beams polarization degree $P_T$.
Therefore a minimal $P_T$ required to achieve $\max_{\theta}\mathcal{F}_3[\rho^{P_T}_{Y\bar{Y}}] \geq 1$ across full $\theta$ range, labeled as $\min P_T$, can be defined as a function of $\phi$, as shown in right panel of Fig.~\ref{fig:FPT2}.
The $\min P_T$ increases as the increase of $\phi \in [0,\pi/2]$ for $\Lambda\bar{\Lambda}$, $\Xi^{-}\bar{\Xi^{-}}$ and $\Xi^{0}\bar{\Xi^{0}}$ pairs, opposite to the trends of $\Sigma^{+}\bar{\Sigma^{+}}$ pair. In particular, for $\Sigma^{0}\bar{\Sigma^{0}}$ pair, the dependence of $\min P_T$ on $\phi$ nearly vanishes, since the critical angle $\theta^*$ is already close to zero in the absence of beam polarization as can also be seen in right panel of Fig.~\ref{fig:maxF_PL}.

\section{Discussions}
\label{sec:discussion}


\subsection{Hierarchy of Quantum Correlations}
\label{sec:hierarchy}

Among four different types of entanglement measures in
quantum information theory characterizing various aspects of quantum correlations of bipartite systems,
a hierarchy of quantumness is expected \cite{Wiseman:2007hyt},
\begin{equation}
\mathrm{Bell\ Nonlocality}\subset\mathrm{Steering}\subset\mathrm{Entanglement}\subset\mathrm{Discord}.\label{eq:hierarchy}
\end{equation}
Here we investigate this hierarchy by facilitating the concept of entanglement of formation (EoF).
EoF is widely utilized as a measure of entanglement for bipartite state $\rho$, and particularly useful for mixed states.
For states of two qubits, the EoF has a close relationship with Wotters' concurrence $\mathcal{C}[\rho]$ \cite{Wootters:1997id}.
\begin{equation}
\mathscr{E}[\rho] = {h} \left[\frac{1}{2}\left({1+\sqrt{1-\mathcal{C}^2[\rho]}}\right)\right].\label{eq:eof}
\end{equation}
which is zero for separable states and non-zero for entangled states.
On the other hand, concurrence places bounds on the CHSH and CJWR parameters as \cite{Verstraete:2001skx,Fan:2020egb,Fan:2022sif}
\begin{align}
  &\mathcal{B}[\rho] \leq 2\sqrt{1+\mathcal{C}^{2}[\rho]},\label{eq:BvsC} \\
  &\mathcal{F}_{3}[\rho] \leq\sqrt{1+2\mathcal{C}^{2}[\rho]}. \label{eq:FvsC}
\end{align}
The CHSH parameter, which quantifies Bell nonlocality \cite{Clauser:1969ny}, has been studied for hyperon pairs produced in  with unpolarized and polarized lepton beams \cite{Wu:2024asu,Zhang:2026nwm}.
A sufficient (but not necessary) condition for Eqs.~(\ref{eq:BvsC}, \ref{eq:FvsC}) to hold as equalities (except for pure state) is that the 2-qubit system is Bell diagonal state with vanishing hyperon polarization \cite{Verstraete:2001skx,Fan:2020egb,Fan:2022sif}.
Then two lower bounds for concurrence can be established as
\begin{align}
&\sqrt{\max \left\{ 0 ,{\frac{1}{4}\mathcal{B}^2[\rho]-1}\right\}} \leq \mathcal{C}[\rho],\\
&\sqrt{\max \left\{0,{\frac{1}{2}({\mathcal{F}_3}^2[\rho]-1)}\right\}} \leq \mathcal{C}[\rho].
\end{align}
Following the definition of EoF we thus define $\mathscr{B}$ and $\mathscr{F}$ as the measures of Bell nonlocality and quantum steering, respectively:
\begin{align}
\mathscr{B}[\rho] &:= h\left[\frac{1}{2}\left(1+\sqrt{1-\max \left\{ 0 , \frac{1}{4}\mathcal{B}^2[\rho]-1\right\}}\right)\right],\\
\mathscr{F}[\rho] &:= h\left[\frac{1}{2}\left(1+\sqrt{1-\max \left\{0,\frac{1}{2}({\mathcal{F}_3}^2[\rho]-1)\right\}}\right)\right].
\end{align}
The measures $\mathscr{B}$ and $\mathscr{F}$ defined in this manner satisfy that:
\begin{itemize}
  \item[i.]  $\mathscr{B},\ \mathscr{F}\in[0,1]$, a non-zero value occurs iff (if and only if) the CHSH or CJWR inequality is violated, and $\mathscr{B}=1$ or $\mathscr{F}=1$ signifies the maximal violation;
  \item[ii.] Eqs.~\eqref{eq:BvsC} and \eqref{eq:FvsC} are reduced to  very simple relations $\mathscr{B}\leq\mathscr{E}$ and $\mathscr{F}\leq\mathscr{E}$, respectively.
\end{itemize}
These properties make them well-suited for comparing four different types of quantum correlations, {which quantify distinct aspects of the non-classicality of mixed states. Quantum discord is the most robust, capturing non-classical correlations even when entanglement is weak. EoF quantifies the non-separability of the state, though not all entangled mixed states exhibit non-local features. Quantum Steering represents a stronger correlation than entanglement, signifying the ability to non-locally influence one hyperon's state through measurements on the other—a phenomenon that cannot be explained by LHS models. Bell nonlocality is strongest among those, manifesting in correlations that violate Bell inequalities and exclude local hidden-variable theories (LHVTs).}

\begin{figure}[!t]
\begin{minipage}{0.41\linewidth}
		\vspace{3pt}
		\centerline{\includegraphics[width=\textwidth]{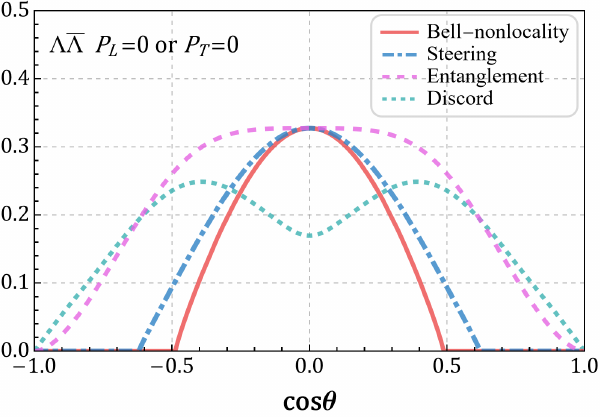}}

	\end{minipage}
	\begin{minipage}{0.285\linewidth}
		\vspace{3pt}
		\centerline{\includegraphics[width=\textwidth]{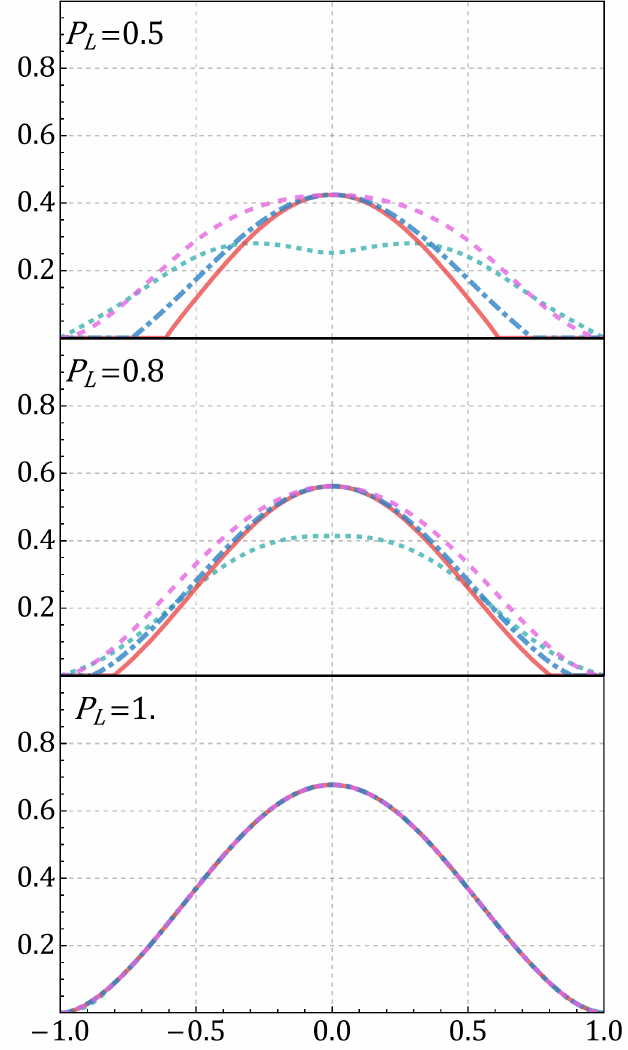}}
	
	\end{minipage}
	\begin{minipage}{0.285\linewidth}
		\vspace{3pt}
		\centerline{\includegraphics[width=\textwidth]{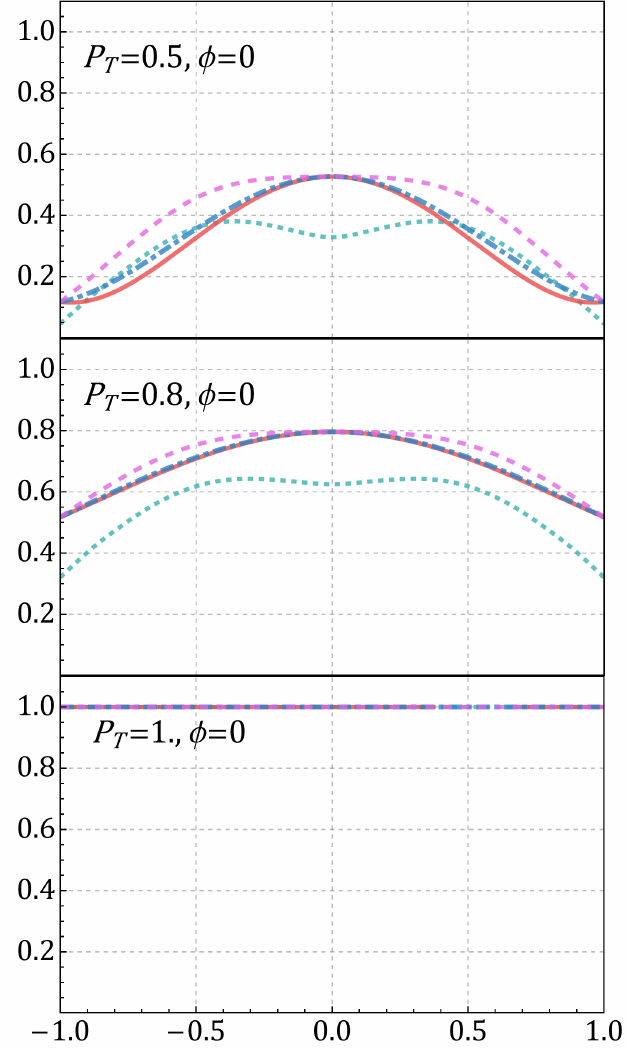}}

	\end{minipage}

\caption{\raggedright
The quantum correlation measures $\mathscr{D},\mathscr{E},\mathscr{F},\mathscr{B}$ for $J/\psi\to \Lambda\bar{\Lambda}$ are compared as functions of $\cos\theta$. left panel: unpolarized lepton beams. right left panels: longitudinal polarized electron beam. right right panels: transversely polarized lepton beams with $\phi=0$. The polarization degrees $P_L$ or $P_T$ are indicated in each panel.
} \label{fig:hierarchy}
\end{figure}

By combining the results in Sec.~\ref{sec:PL}, \ref{sec:PT} and our earlier work \cite{Zhang:2026nwm}, the  relation $\mathscr{B}=\mathscr{F}=\mathscr{E}$ is always held at $\theta=0,\pi\text{ and }{\pi}/{2}$ for hyperon-antihyperon pairs regardless of the polarization of beams.
A sufficient condition for the coincidence of the four types of quantum correlations is that the system is in a pure state \cite{Han:2024ugl,Verstraete:2001skx,Fan:2020egb,Fan:2022sif}.
When $P_L=1$ or $P_T=1$, the hyperon pair system is in a pure state and one can prove that
\begin{equation}
  \begin{aligned} &\mathscr{B}=\mathscr{F}=\mathscr{E}=\mathscr{D}=S\left(\rho_{Y(\bar{Y})}\right)=h\left[\frac{1}{2}\left(1+\left\|\vec{B}\right\|\right)\right], \label{eq:equality}
  \end{aligned}
\end{equation}
with $\left\|\vec{B}\right\|$ given in Eqs. \eqref{eq:rdmeigenPL} and  \eqref{eq:rdmeigenPT} by setting $P_L=1$ and $P_T=1$, respectively.
{The hierarchical structure stems from the distinct interpretations of quantumness by different quantum correlations in mixed-state systems admixed with classical noise. In the pure-state limit, the absence of classical mixing implies that all non-classical measures reduce to the entanglement entropy, i.e., the local von~Neumann entropy.}


Taking $\Lambda\bar{\Lambda}$ as an example, Fig.~\ref{fig:hierarchy} compares the evolution of different measure of quantum correlations as the longitudinal or transverse beams polarization degree increases.
For the case of transverse beam polarization a fixed angle $\phi = 0$ is shown.
When $P_L$ or $P_T$ approaches the maximum degree, and four quantum correlations become close to each other and ultimate equal at any scattering angle when the system is in a pure state (see Eq.~\eqref{eq:equality}).
{The beam polarization determines the purity of the system. As the polarization degree drops, stronger correlations such as steering and Bell nonlocality decrease more rapidly, as they are highly sensitive to the state mixing introduced by lowering the beam polarization degree. Physically, this transition acts as a decoherence-like mechanism; however, it originates from a statistical mixture rooted in the classical uncertainty of the initial state preparation rather than dynamical environment-induced decoherence. These behaviors reinforce the motivation that hyperon systems provide a unique platform for understanding the hierarchy of quantum correlations if polarized beams are implemented.}
It is also evident that the magnitudes of Bell nonlocality, steering, and entanglement (quantified by the EoF) strictly adhere to the ordering $\mathscr{B}\leq \mathscr{F}\leq \mathscr{E}$ given in Eq.~\eqref{eq:hierarchy}, regardless of $P_L$ or $P_T$.
In contrast, the quantum discord $\mathscr{D}$ does not obey this hierarchy; in many cases, its value is even lower than that of the Bell nonlocality.
Consequently, for mixed states, quantum discord cannot be used to directly infer its position within the hierarchy of the four quantum correlations \cite{Luo:2008ecu,Wu:2025dds}.
We address this issue further in the next subsection \ref{sec:beyond} by analyzing separable states.

\subsection{Beyond Entanglement: Quantum Discord} \label{sec:beyond}


A system's state is described by corresponding spin density matrix $\rho$ acting on its Hilbert space ${\mathcal{H}}$.
If one describes $Alice$’s physical system by the Hilbert space ${\mathcal{H}_A}$ and that of $Bob$ by ${\mathcal{H}_B}$, their joint physical system is described by the tensor product Hilbert space ${\mathcal{H}_A} \otimes {\mathcal{H}_B}$.
A mixed state $\rho_{AB}$ is separable if and only if it can be represented as a convex combination of the product of density operators on local states \cite{Werner:1989zz}:
\begin{equation}
\rho_{AB} = \sum_i c_i \rho_{A}^i \otimes \rho_{B}^i,
\label{eq:Sep_state}
\end{equation}
where $\{c_i\}$ are probabilities satisfying $\sum_i c_i=1$.
Otherwise $\rho_{AB}$ is entangled.

In quantum information theory, the set of zero-discord states forms a strict subset of separable states \cite{Han:2024ugl}.
For a bipartite system, separable state can be classified as locally classical if the state remains invariant under local measurements on one subsystem; otherwise, it is considered to exhibit quantum correlations~\cite{Luo:2008kvk,Modi:2012baj}.
Depending on which subsystem is considered, this criterion leads to a classification of bipartite states:
quantum-quantum (q-q), quantum-classical (q-c), classical-quantum (c-q), and classical-classical (c-c).
Since the $CP$ violation is not considered here, the polarization vector and spin correlation tensor of the $Y\bar{Y}$ system are symmetric, implying that the states $\rho_{Y\bar{Y}}$ can only be either q-q or c-c \cite{Han:2024ugl}.
Zero-discord states can be  characterized as either q-c or c-q states, with c-c states constituting a specific subset of both. Therefore, the zero-discord states of the $Y\bar{Y}$ system must necessarily be of the c-c type; otherwise, they fall into the q-q category.

\begin{figure}[!t]
\centering
  \includegraphics[width = 0.495 \linewidth]{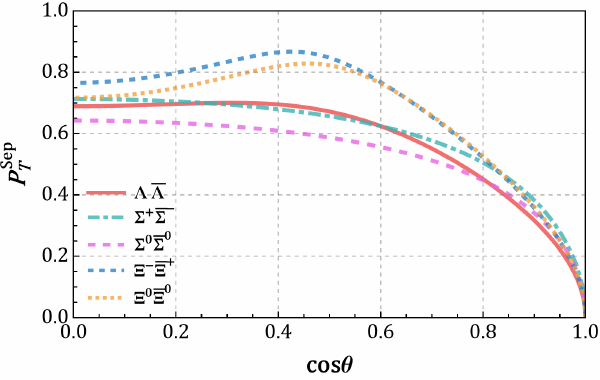}
  \includegraphics[width = 0.495 \linewidth]{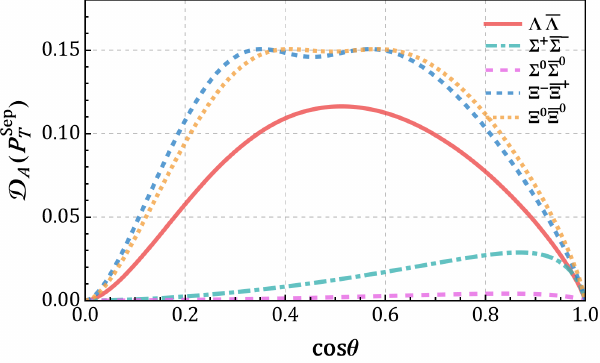}
\caption{\raggedright
The $P_T^{Sep}$ (left panel) and $\mathscr{D}_A(P_T^{Sep})$ (right panel) as a function of $\cos\theta$ in $J/\psi\to Y{\bar{Y}}$ for $Y=\Lambda$, $\Sigma^+$, $\Sigma^0$, $\Xi^{-}$ and $\Xi^{0}$. See main text for detailed explanation.
} \label{fig:SepD}
\end{figure}

In our previous work, we identified the region within which the entanglement measured by concurrence and Bell nonlocality of $Y\bar{Y}$ system are vanishing under the transverse beam polarization \cite{Zhang:2026nwm}.
This occurs at certain azimuthal angles of
\begin{itemize}
  \item[a.] $\phi^* = k\pi,(k=0,1,2)$ for $\Sigma^{+}\bar{\Sigma}^{-}$, $\Sigma^{0}\bar{\Sigma}^{0}$ pairs with $\alpha_\psi<0$;
  \item[b.] $\phi^* = \pi/2+k\pi,(k=0,1)$ for $\Lambda\bar{\Lambda}$, $\Xi^{-}\bar{\Xi}^{+}$, and $\Xi^{0}\bar{\Xi}^{0}$ pairs with $\alpha_\psi>0$.
\end{itemize}
which is also applicable to steering in Sec. \ref{sec:PT}.
For a fixed scattering angle $\theta$, separable states $\rho^{Sep}_{Y\bar{Y}}$ can exist for a specific beam polarization degree denoted as $P_T^{Sep}$. As shown in left panel of Fig.~\ref{fig:SepD}, the curve characterize the $P_T^{Sep}$ as a function of $\cos \theta$ for different hyperon pairs. It can be seen that separable state occurs at $\theta=0,\pi$ iff the beams are unpolarized.

The discord of separable states, denoted as the $\mathscr{D}_{A}(P_T^{Sep})$, are showed as a function of $\cos\theta$ in right panel of Fig.~\ref{fig:SepD}.
The $\mathscr{D}_{A}(P_T^{Sep})$ increases to a maximum and then decreases as the increasing of $\cos\theta$ for $\Lambda\bar{\Lambda}$, $\Sigma^{+}\bar{\Sigma}^{-}$ and $\Sigma^{0}\bar{\Sigma}^{0}$ system.
It has two extreme values within $\cos \theta \in [0,1]$ for $\Xi^{-} \bar{\Xi^{-}}$ and $\Xi^{0} \bar{\Xi^{0}}$ system.
$\mathscr{D}_{A}(P_T^{Sep})$ is vanishing at $\cos\theta=0,\pm1$ for all pairs,
whereas in other regions it remains non-zero even though the system is separable.
For $\Sigma^{+}\bar{\Sigma}^{-}$ and $\Sigma^{0}\bar{\Sigma}^{0}$ system $\mathscr{D}_{A}(P_T^{Sep})$ is relatively small, indicating that quantumness of those hyperon-anti-hyperon system are scarce when they are in separable states.
The result here demonstrates discord as a form of quantum correlation beyond entanglement.

A computable discord witness for a bipartite system $\rho$ is that if and only if the $\mathrm{rank}(\vec{C})>d_A$ with $d_A=\mathrm{dim}(\mathcal{H}_{A})$ and $\vec{C}$ being the correlation matrix of $\rho$, then the state has a non-zero discord $\mathscr{D}_A$\cite{Dakic:2010xfz}.
Given that $d_A=2$ and \(\vec{C}\) is a \(3 \times 3\) real matrix for 2-qubit system here, the discord is non-vanishing iff \(\vec{C}\) is of full rank.
Considering $\vec{C}$ is of full rank iff $\det({\vec{C}}) \neq 0$, the condition $\det({\vec{C}}) = 0$ spans the set of zero discord states.
For hyperon system in the case of transverse beam polarization, the correlation matrix $\vec{C}_T$ in Eq,~\eqref{eq:CT} at $\phi^*$ is
\begin{align}
\vec{C}_T(\phi^*)=\frac{1}{\chi_T}
\begin{pmatrix}
C_{xx} & 0 & C_{xz}\\
0 & C_{yy} & 0\\
 C_{zx}& 0 &C_{zz}
\end{pmatrix}
\end{align}
where the non-zero elements at $\phi^*$ are
\begin{equation}
  \begin{aligned}
    &C_{xx}=\sin^2\theta+{P_T^2}(\alpha_{\psi}\pm\cos^2\theta),\\
    &C_{yy}=-\alpha_{\psi}\sin^2\theta\mp{P_T^2} (1+\alpha_{\psi}\cos^2\theta),\\
    &C_{zz}=\alpha_{\psi}\pm\cos^2\theta+{P_T^2}\sin^2\theta,\\
    &C_{xz}=C_{zx}=\gamma_{\psi}\sin\theta\cos\theta(1\pm{P_T^2}).
  \end{aligned}
\end{equation}
The upper and lower signs of ``$\pm$'' and ``$\mp$'' are for case-$a$ and case-$b$ of $\phi^*$, respectively.
In general, the condition $\det\left({\vec{C}_T(\phi^*)}\right) \neq 0$ is satisfied whenever $\rho^{Sep}_{Y\bar{Y}}$ represents a q-q state. The only exceptions arise at $\theta=0, \pi$ and $\theta=\pi/2$, where the state $\rho^{Sep}_{Y\bar{Y}}$ reduces to a c-c state, as detailed below.

At $\theta=0,\pi$, the correlation matrix $\vec{C}_T$ reduces to be diagonal:
\begin{align}
  \frac{1}{1+\alpha_{\psi}}
\begin{pmatrix}
{P_T^2}(1\pm\alpha_{\psi}) & 0 & 0 \\
0 & \mp{P_T^2}(1+\alpha_{\psi}) & 0\\
0 & 0 & 1\pm\alpha_{\psi}
\end{pmatrix}.
\end{align}
which is of full rank if \( \alpha_{\psi} \neq \pm 1 \) and \( P_T \neq 0 \),
thus transverse beam polarization induces non-zero discord of hyperon system at $\theta=0,\pi$ and $\phi^*$.
For the case of \( \alpha_{\psi} = \pm 1 \) corresponding to structureless elementary particles \cite{Fang:2026ddi}, the correlation matrix \( \vec{C}_T(\phi^*) \) is rank-deficient, and tuning the transverse beam polarization can not generate discord of the system at those specific angles.

At $\theta=\pi/2$, the correlation matrix $\vec{C}_T(\phi^*)$ reduces to
\begin{align}
 \frac{1}{ 1\pm{P_T^2}\alpha_{\psi}}
\begin{pmatrix}
1+{P_T^2}\alpha_{\psi}& 0 & 0 \\
0 & -\alpha_{\psi}\mp{P_T^2} & 0\\
0 & 0 & \alpha_{\psi}+P_T^2
\end{pmatrix}.
\end{align}
which is rank-deficient iff $\alpha_{\psi}=\pm{P_T^2}$.
This condition determines $P_T^{Sep}(\cos\theta=0)=\sqrt{|\alpha_{\psi}|}$ as shown in the left panel of Fig.~\ref{fig:SepD}.
For elementary particles with \( \alpha_{\psi} = \pm 1 \), maximizing the transverse beam polarization completely eliminates the quantum nature of the system at those specific angles \cite{Fang:2026ddi}.

\subsection{Experimental Sensitivity} \label{sec:exp}

\begin{figure}[!htbp]
\centering
  		\includegraphics[width = 0.7\linewidth]{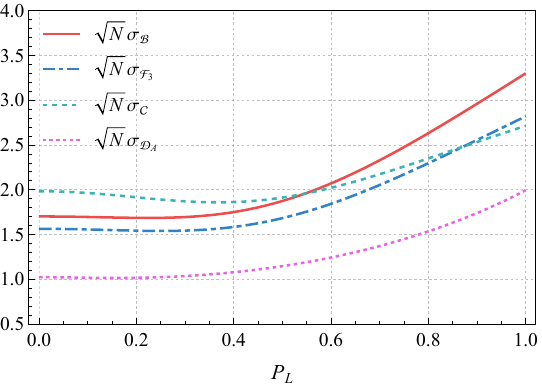}
\caption{\raggedright The statistical uncertainty of $\mathscr{D}_{A}$, $\mathcal{F}_3$, Bell nonlocality $\mathcal{B}$ and concurrence $\mathcal{C}$ scaled by $\sqrt{N}$ at $\theta = \pi/2$ as a function of the longitudinal polarization degree for $J/\psi\to \Lambda (p \pi^-)\bar{\Lambda} (\bar{p} \pi^+)$.
} \label{fig:SUPL}
\end{figure}

\begin{figure}[!htbp]
\centering
  		\includegraphics[width = 0.4\linewidth]{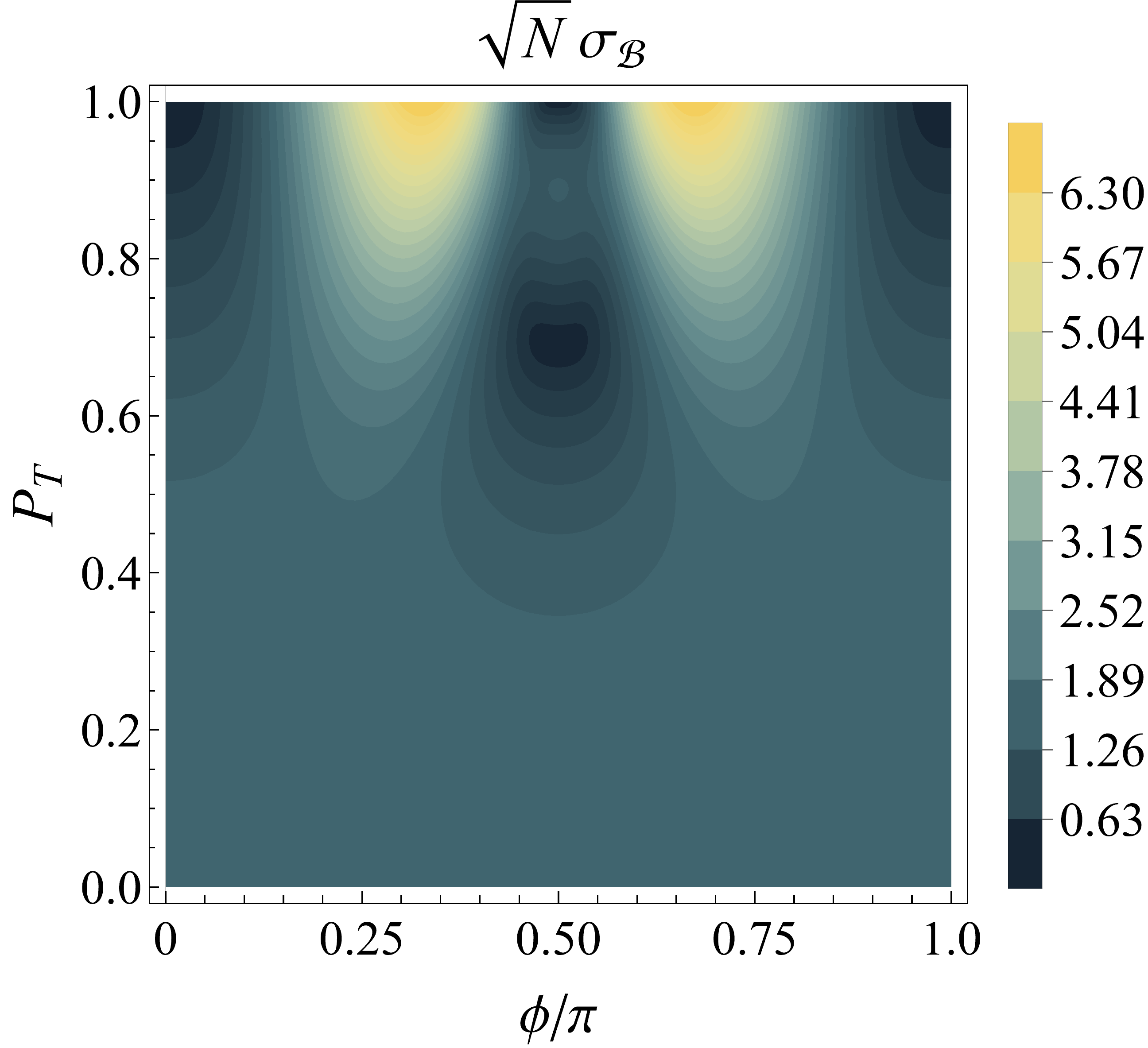}
        \includegraphics[width = 0.4\linewidth]{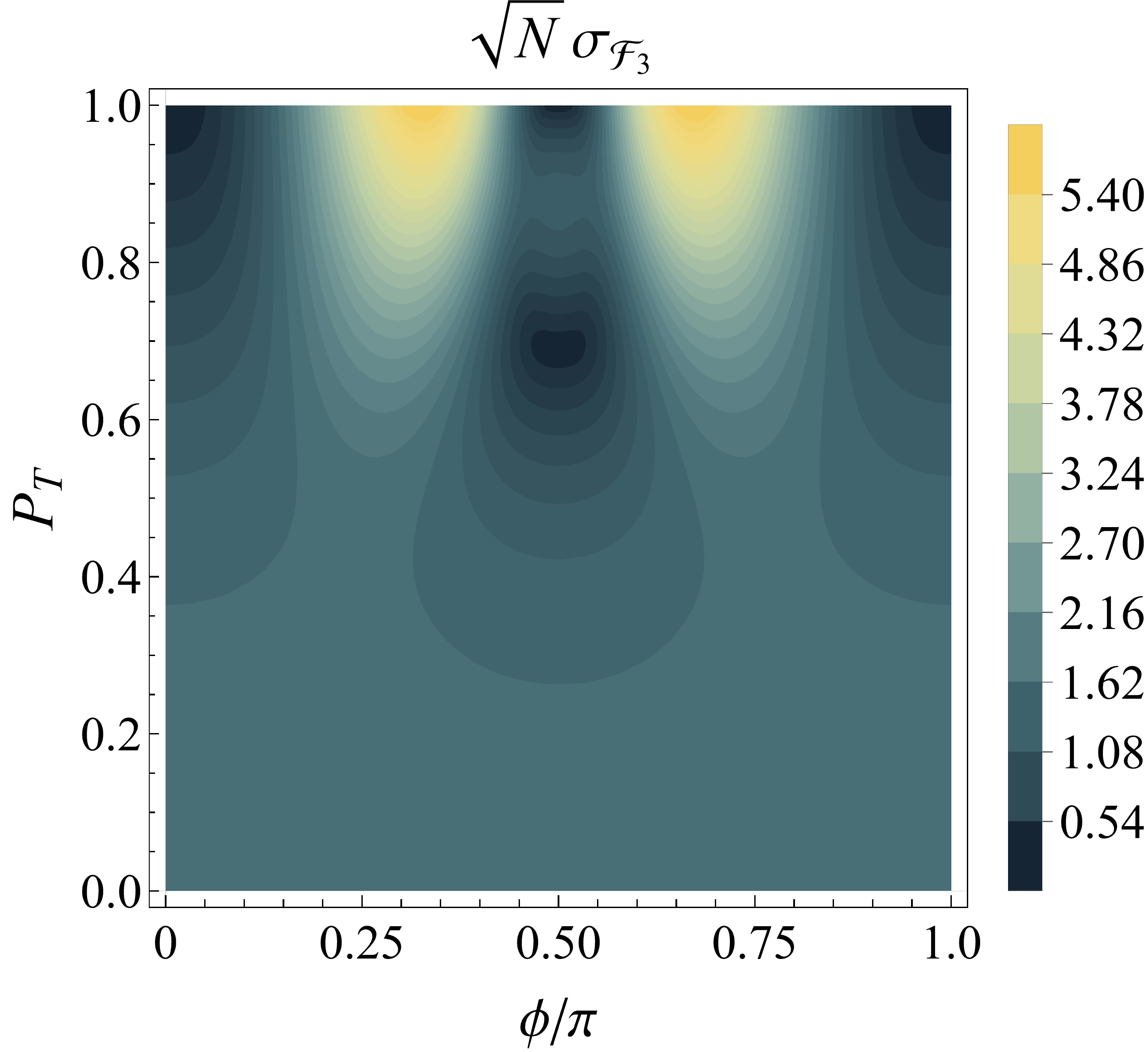}
        \includegraphics[width = 0.4\linewidth]{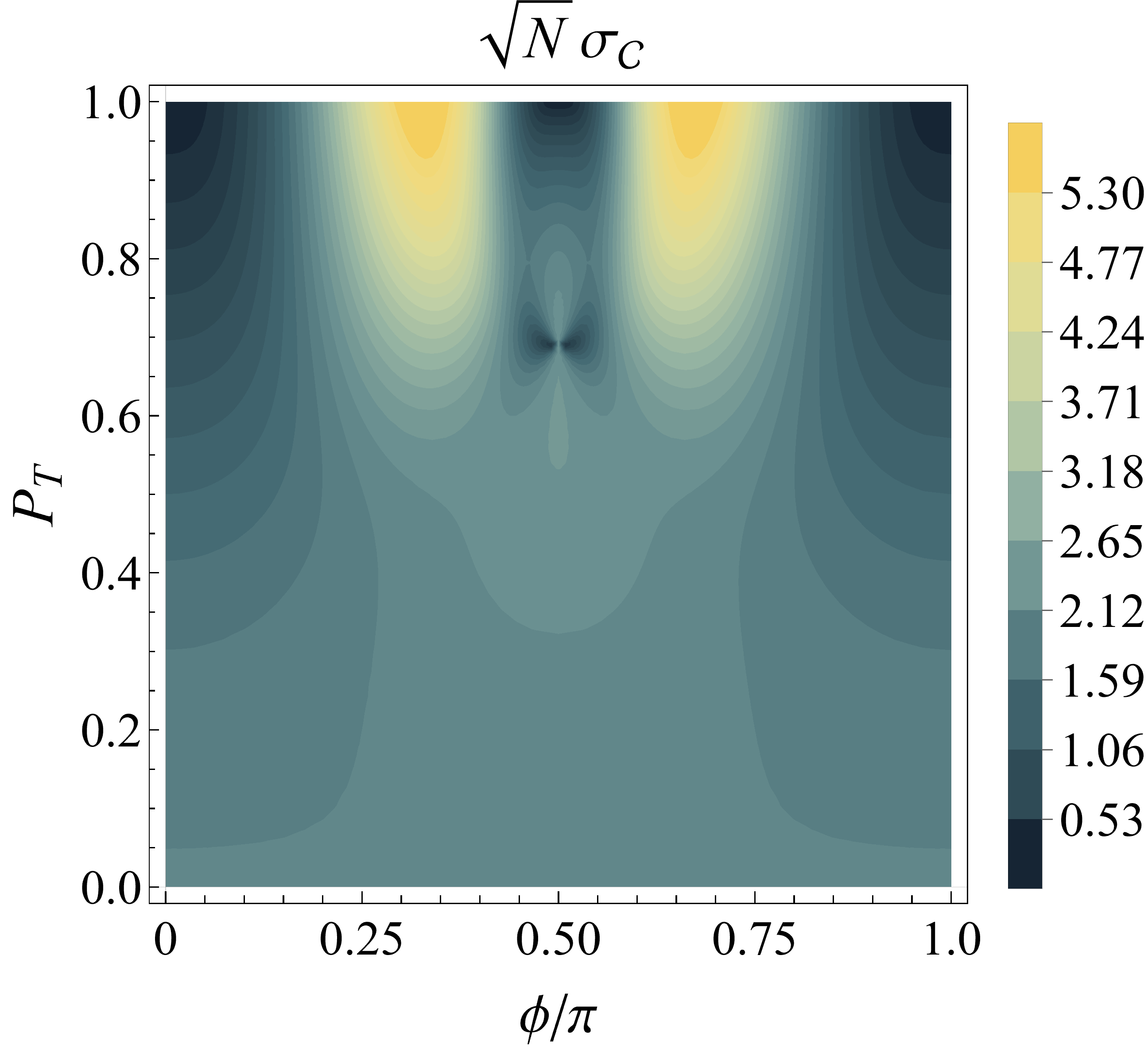}
        \includegraphics[width = 0.4\linewidth]{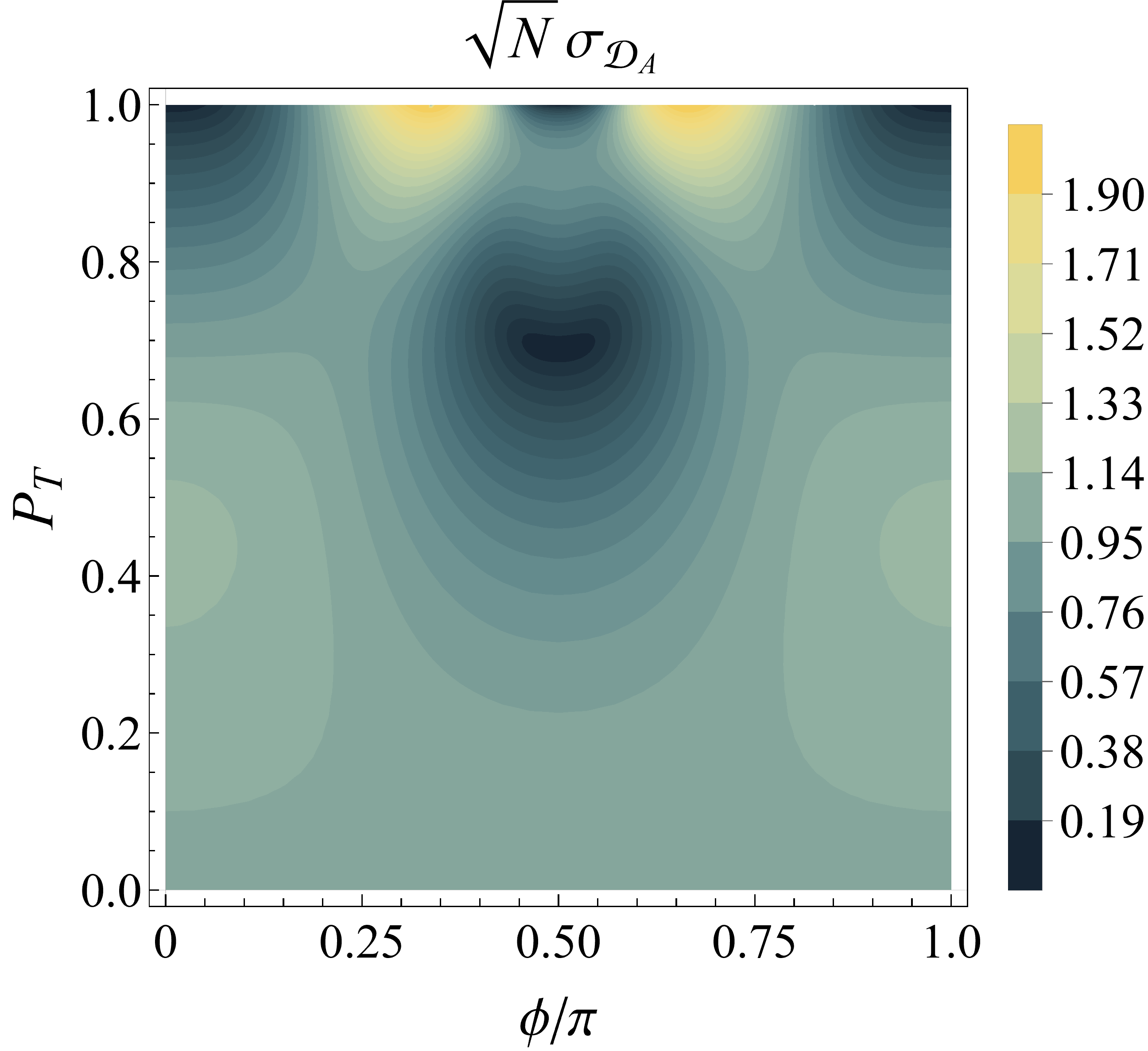}
\caption{\raggedright The statistical uncertainty of $\mathscr{D}_{A}$, $\mathcal{F}_3$, Bell nonlocality $\mathcal{B}$ and concurrence $\mathcal{C}$ scaled by $\sqrt{N}$ at $\theta = \pi/2$ as a function of the transverse polarization degree for $J/\psi\to \Lambda (p \pi^-)\bar{\Lambda} (\bar{p} \pi^+)$
} \label{fig:SUPT}
\end{figure}

{
Assessing the experimental sensitivity through toy Monte Carlo simulations would be highly valuable to clarify the significance of quantum discord and steering in actual data analysis. However, implementing such simulations at this stage presents a significant computational bottleneck for quantum discord evaluation.
As noted in Eq.~\ref{eq:POVM}, computing quantum discord requires an optimization of classical mutual information over all local von Neumann measurements on one subsystem. This process is computationally intensive and currently too slow to be performed for large simulated event samples. Significant code optimization or parallelization would be required to overcome this---a challenge that will also affect future experimental data analyses involving discord. While analytical expressions for quantum discord have been established for simpler elementary particle systems (e.g., top-antitop in \cite{Han:2024ugl}), they are currently unavailable for baryon systems. Addressing this remains a non-trivial challenge for future study.
Therefore, a simple estimation of experimental sensitivity based on the expected statistical uncertainty is performed for quantum steering and discord at the specific scattering angle $\theta = {\pi}/{2}$, as analytical expressions can be derived at this angle, allowing for a more direct assessment of the experimental sensitivity within the maximum likelihood estimation \cite{Salone:2022lpt}.

If only one of the initial beams is considered to be longitudinally polarized, the use of longitudinally polarized beam does not modify the effective production cross section \cite{Salone:2022lpt,Zhao:2025cbd}; therefore, the available statistics remain unchanged. The same conclusion applies to the case of transversely polarized beams \cite{ Cao:2024tvz}.
Therefore, the statistical significance of the parameters is evaluated using the log-likelihood derived from the joint angular distribution of $J/\psi\to \Lambda (p \pi^-)\bar{\Lambda} (\bar{p} \pi^+)$, as explicitly formulated in the literature:
\begin{align}
\sigma^{-1}(\alpha_\psi) &= \sqrt{\frac{N}{5} \big(1+ \frac{23}{27} \alpha_\Lambda^4 + \frac{10}{9} \alpha_\Lambda^2 P_L^2\big)},  \quad  \sigma^{-1}(\beta_\psi) = \alpha_\Lambda \sqrt{\frac{4N}{45} \big( 1 + \frac{5}{3} \alpha_\Lambda^2 P_L^2 \big)}, \nn \\  \sigma^{-1}(\gamma_\psi) & = \alpha_\Lambda^2 \sqrt{ \frac{4 N}{135} \big(\alpha_\Lambda^2 +15 P_L^2 \big)}. \nn 
\end{align}
for longitudinally polarized beam \cite{Salone:2022lpt}, and
\begin{align}
\sigma^{-1}(\alpha_\psi) & = \sqrt{\frac{N}{5} \big(1 + \frac{23}{27} \alpha_\Lambda^4 +\frac{4}{3} P_T^4 \big)}, \quad  \sigma^{-1}(\beta_\psi) = \alpha_\Lambda \sqrt{\frac{4 N}{45} (1 + 3 P_T^4)}, \nn \\ \sigma^{-1}(\gamma_\psi)  &= \alpha_\Lambda^2 \sqrt{\frac{4 N}{135} (1 + 3 P_T^4)}. \nn 
\end{align}
for transversely polarized beams \cite{Cao:2024tvz}.
Here $N$ is the reconstructed event yield. CP conservation is imposed by taking $\alpha_\Lambda = -\alpha_{\bar{\Lambda}} = 0.75$ \cite{ParticleDataGroup:2026mpi}.
The correlations between parameters and errors of decay parameters $\alpha_\Lambda$ of $\Lambda \to p \pi^-$ are both neglected.
As can be seen polarized beams enhance the sensitivity for parameters $\alpha_\psi$, $\beta_\psi$ and $\gamma_\psi$ (and thus $R_{EM}$ and $\Delta\Phi$).

Neglecting detector acceptance and coverage, standard error propagation is used to estimate the statistical uncertainties of $\mathscr{D}_{A}$ and $\mathcal{F}_3$ at $\theta = \pi/2$, utilizing the analytical expressions derived in the preceding sections, together with Bell nonlocality $\mathcal{B}$ and concurrence $\mathcal{C}$ in our previous work \cite{Zhang:2026nwm}.
Figs.~\ref{fig:SUPL} and ~\ref{fig:SUPT} illustrate the statistical uncertainty scaled by $\sqrt{N}$ as a function of the polarization degree for all entanglement measures. It is evident that the statistical uncertainties increase only mildly with the polarization degree. Notably, achieving a precision of $0.01$ requires at most 0.25 million observed events per bin. Given the total sample of 3.2 million events collected by the BESIII experiment \cite{BESIII:2022qax}, the available statistics remain more than sufficient to meet this requirement, even after partitioning the data into multiple bins. By examining the evolution of the magnitudes of $\mathscr{D}_{A}$, $\mathcal{F}_3$, Bell nonlocality and concurrence versus the polarization degree, it is feasible to observe the control of quantum entanglement with statistical significance provided that a similar event yield is achieved under a polarized beam configuration with a value of $P_{L,T} > 0.6$.
Therefore, systematic uncertainties not considered in the present analysis will be of significant importance for future experimental investigations.

As can also be observed, achieving the same statistical uncertainty requires a smaller event sample for $\mathscr{D}_{A}$ than for other three measures, reflecting the hierarchy established in Eq.~\eqref{eq:hierarchy}. Furthermore, the statistical uncertainties of the latter three quantities are very similar to each other. The procedure developed here can be utilized for the rapid estimation of other quantum tomography observables.

The experimental sensitivity analysis presented above focuses on the $J/\psi \to \Lambda\bar{\Lambda}$ channel. For the $J/\psi \to \Sigma^{+}\bar{\Sigma}^{-}$ and $J/\psi \to \Sigma^{0}\bar{\Sigma}^{0}$ channels, the accumulated events at BESIII are approximately 0.73 million and 1.0 million, respectively~\cite{BESIII:2025jxt,BES:2008hwe,BESIII:2020fqg,BESIII:2024nif}.
However, since the entanglement in these channels responds weakly to beam polarization as explained in the preceding sections, it is unfeasible to observe the control of quantum discord and steering in $Sigma$ hyperon channels.
In contrast, evaluating the experimental sensitivity for $J/\psi \to \Xi^{-}\bar{\Xi}^{+}$ and $J/\psi \to \Xi^{0}\bar{\Xi}^{0}$ is more complex. In these processes, the two-step sequential decay of the $\Xi$ hyperon helps improve the precision of the parameters $\alpha_\psi$, $\beta_\psi$, and $\gamma_\psi$. With approximately 0.6 million and 0.3 million events available at BESIII \cite{BESIII:2026hgj,BESIII:2021ypr,BESIII:2016nix,BESIII:2023drj}, it would be possible to observe the modulation of quantum discord and steering if a comparable number of events were accumulated after the implementation of polarized beams.
}

\section{Conclusion} \label{sec:conclusion}

This work systematically investigate how polarized lepton beams control the quantum discord and steering in hyperon pairs produced in electron-positron annihilation.
By exploiting the joint spin density matrix, numerical calculations are performed to examine the dependence of quantum discord and the three-setting CJWR steering parameter on the degree of longitudinal and transverse beam polarization.
It is apparent that increase of beam longitudinal polarization degree $P_L$ widens the angular domain of CJWR steerability, and moves the largest CJWR parameter and discord to scattering angle perpendicular to the beam direction.
It is theoretically observed that the largest CJWR parameter and discord with respect to scattering angle respond monotonically to the longitudinal polarization degree of beam, failing to attain maximal CJWR inequality violation and discord even at maximum polarization degree.
In contrast, transverse polarization enables the system to be maximal steerability and discord across full range of scattering angle if choosing optimally the azimuthal angle.

It is found that the CJWR parameter and discord in $J/\psi\rightarrow \Sigma^{+} \bar{\Sigma^{+}}$ and $\Sigma^{0} \bar{\Sigma^{0}}$ is insensitive to $P_L$, along with the behavior of Bell nonlocality and concurrence as found in our previous work \cite{Zhang:2026nwm}.
The underlying reason is ascribed to the smallness of their relative phase $\Delta\Phi$ between electric and magnetic couplings.
The resulting small $\beta_\psi$ makes the weak dependence of Bell nonlocality, concurrence, and the CJWR parameter on $P_L$.
For discord the reason is the cancellation arising from $\gamma_{\psi} \simeq \sqrt{1-\alpha_{\psi }^2}$.

In order to investigate the hierarchy of different quantum correlations, a comparison in the spirit of entanglement of formation is proposed.
It turns out that entanglement is necessary but not sufficient for steering, while steering is necessary but not sufficient for nonlocality, as claimed in literature.
The justification of the hierarchy of discord is more involved, yet separable states under the case of the transverse beam polarization offer a useful perspective for this matter.
The angular regime is identified where quantum discord survives despite the complete disappearance of entanglement of hyperon system.
This phenomenon occurs at some specific azimuthal angles, where the correlation matrix of the hyperon pair is of full rank, yet non-diagonal correlation is missing.
This directly demonstrates that quantum discord constitutes a more general form of quantum correlation, capturing quantum features that entanglement measures fail to characterize.
Interestingly, transverse beam polarization can eliminate or generate discord within certain angular domains, thereby driving the system into a classical or quantum state.

{An assessment of the experimental sensitivity is provided, thereby strengthening the experimental relevance of our results.}
This study not only contributes to a deeper understanding of the physical nature of quantum correlations in high-energy processes but also provides a theoretical reference for future quantum information experiments at electron-positron colliders.

\begin{acknowledgments}

The authors would like to thank Yutie Liang and Yuxiang Zhao from the BESIII Collaboration for their valuable input on the assessment of experimental sensitivity.
This work is supported by the National Key R\&D Program of China under Grant No. 2023YFA1606703, the National Natural Science Foundation of China  (Grant Nos. 12547111 and 12235008), and the Hebei Natural Science Foundation with Grant Nos. A2022201017 and A2023201041, and Natural Science Foundation of Guangxi Autonomous Region with Grant No. 2022GXNSFDA035068.
\end{acknowledgments}

\appendix

\section{Supplementary figures for \texorpdfstring{$J/\psi\to\Sigma^0\bar{\Sigma}^0$}{}}
\label{sec:Sigma0}

\begin{figure}[!htbp]
\centering
  		\includegraphics[width = 0.35\linewidth]{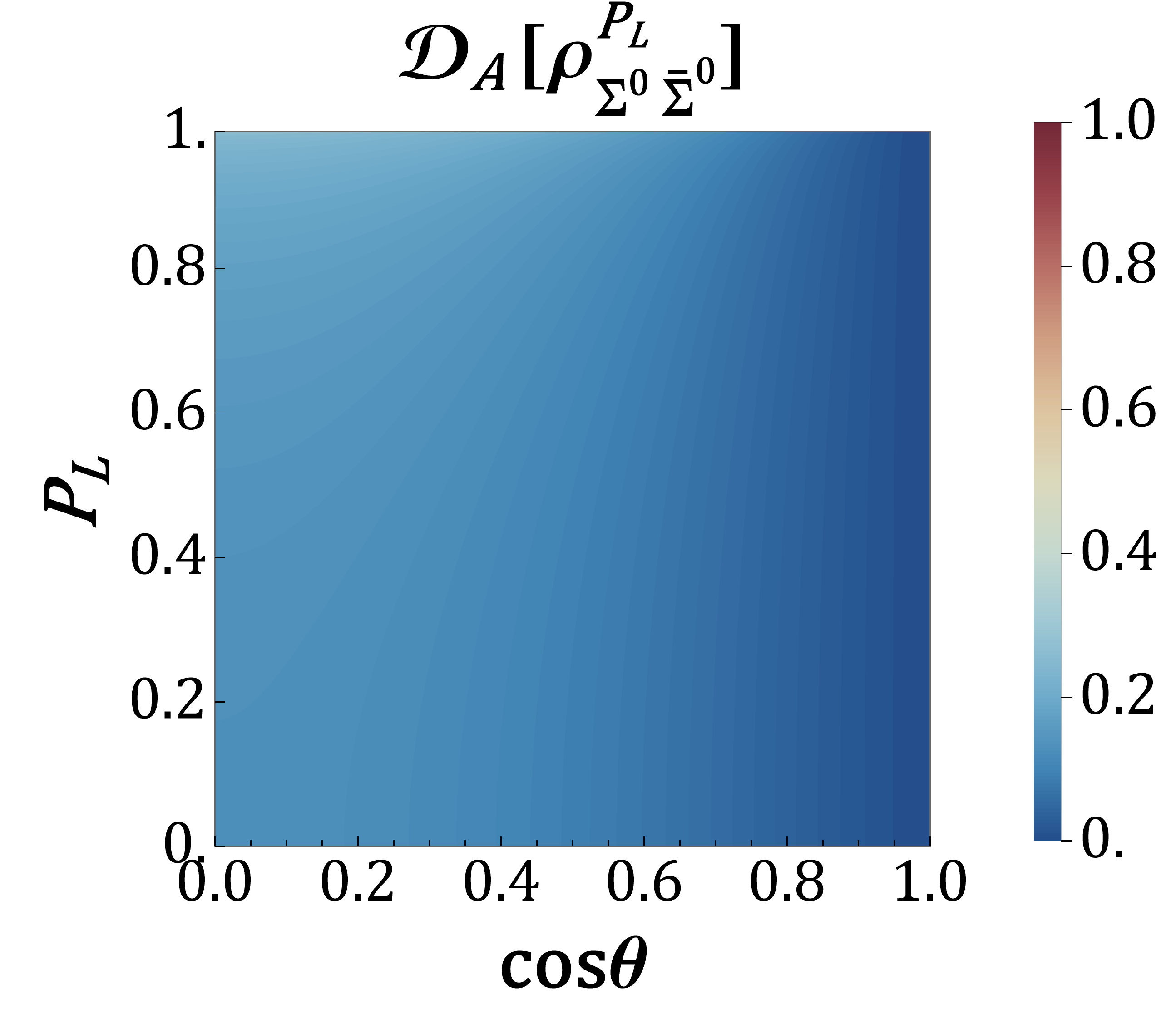}
        \includegraphics[width = 0.35\linewidth]{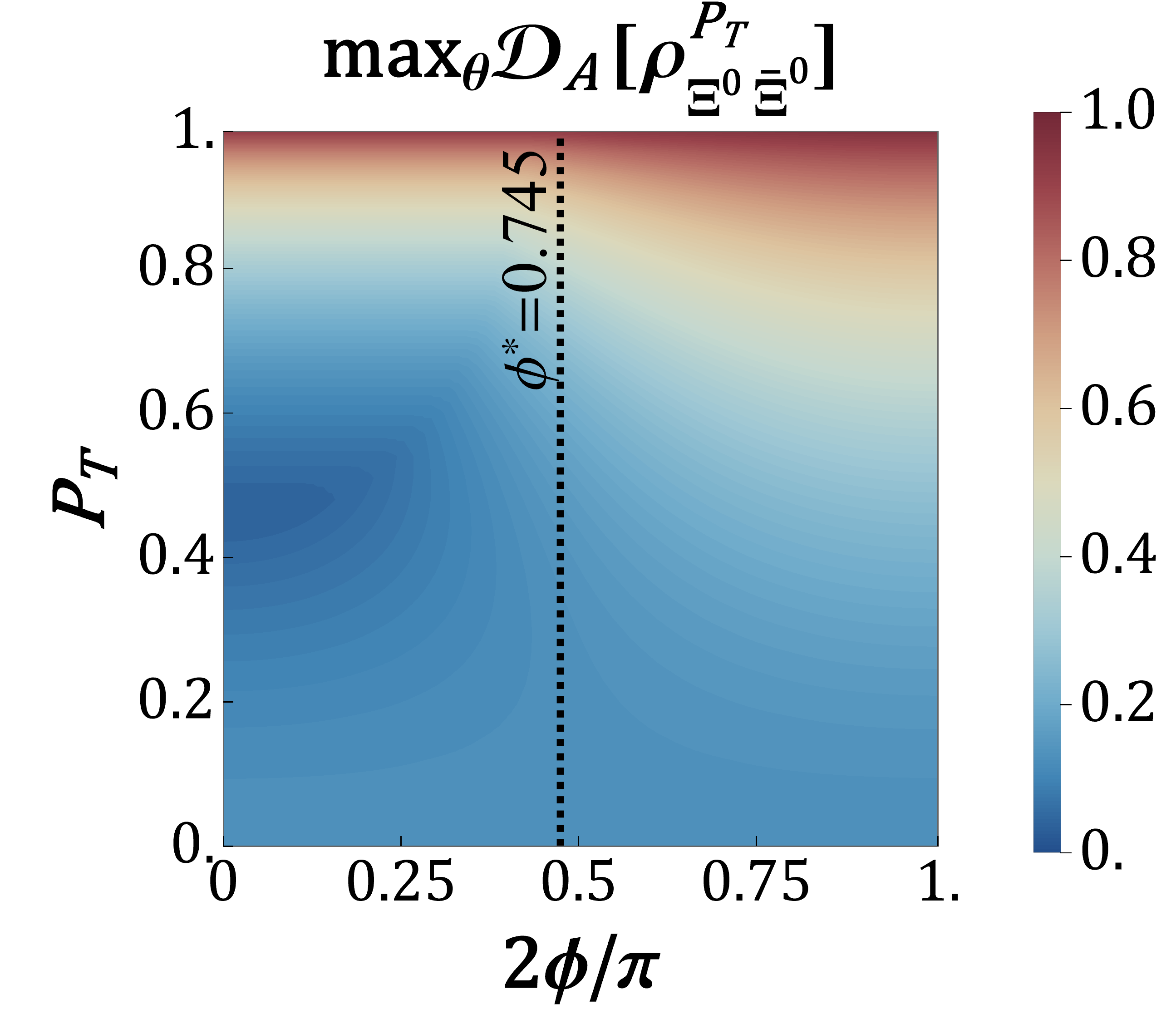}
  		\includegraphics[width = 0.35\linewidth]{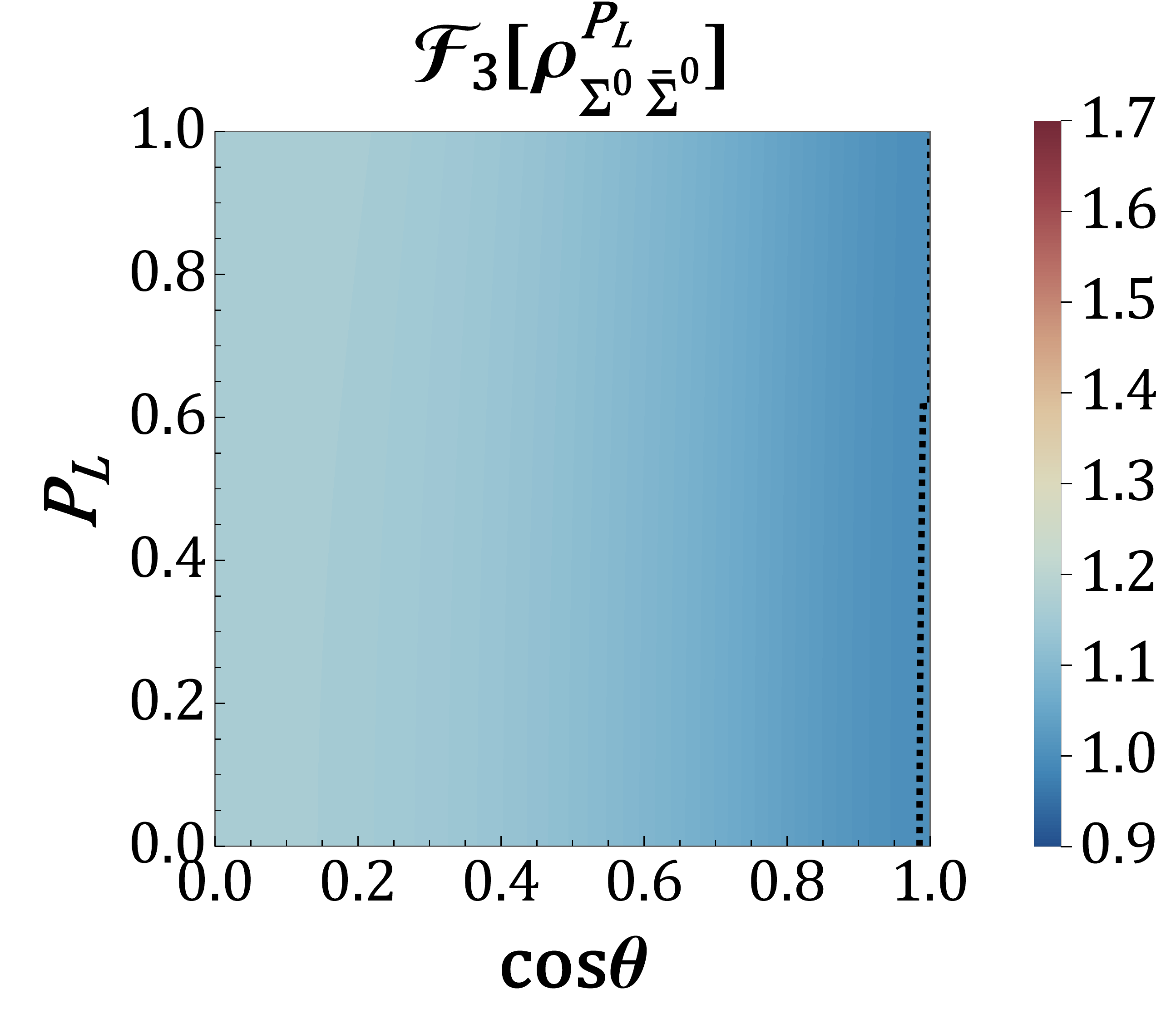}
		  \includegraphics[width = 0.35 \linewidth]{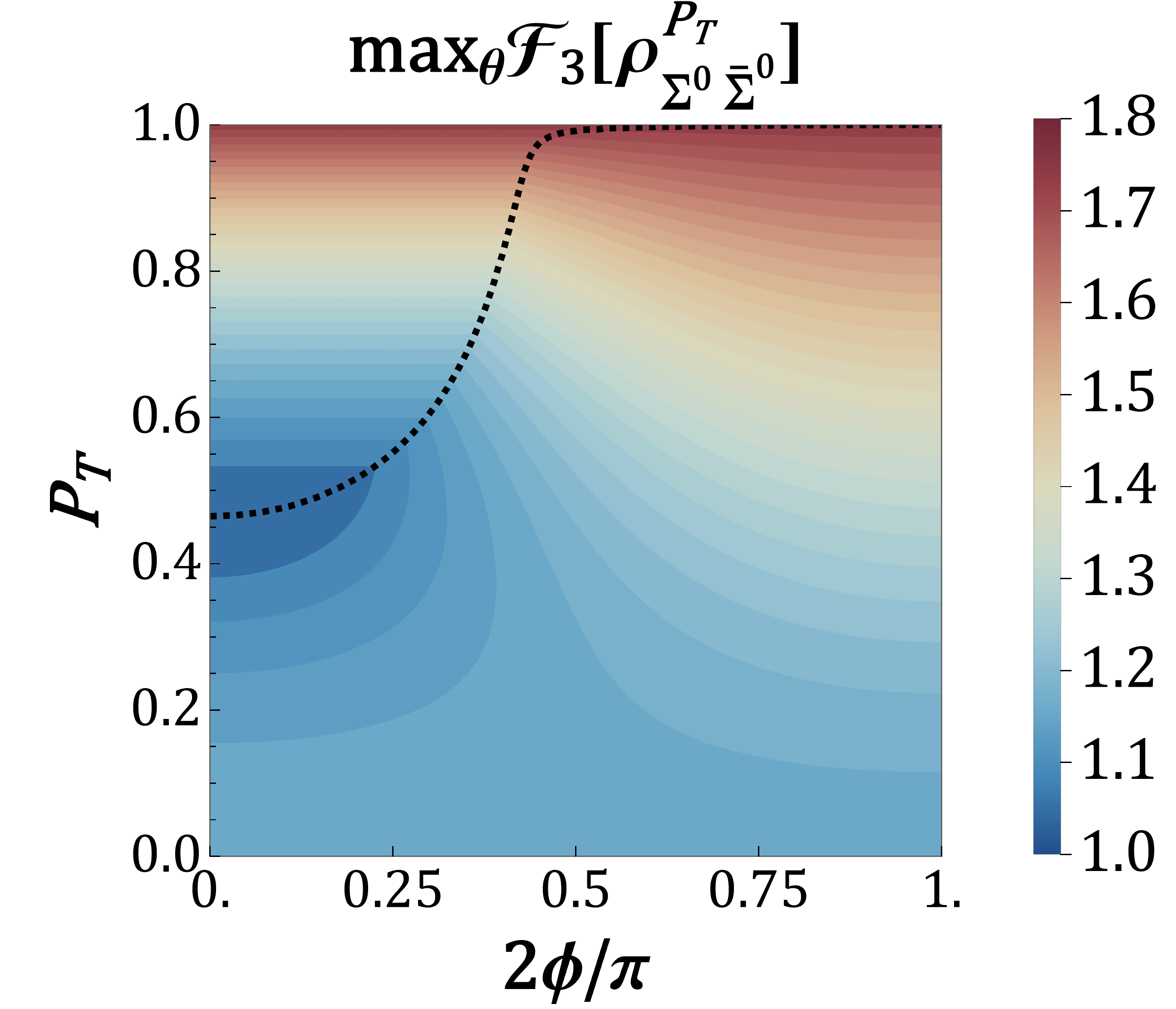}
\caption{\raggedright
The quantum discord and CJWR parameter for $J/\psi \to \Sigma^0{\bar\Sigma}^0$. See main text for the explanation of the dashed curves.} \label{fig:sigma0}
\end{figure}

In Fig.~\ref{fig:sigma0} the quantum discord and the CJWR parameter for $J/\psi \to \Sigma^0{\bar\Sigma}^0$ are shown as a complementary figures.

\section{Eigenvalue decomposition of \texorpdfstring{$\rho_{Y\bar{Y}}^{P_L}$}{}}
\label{sec:eigen}
A pure state of two-qubit system under the $\sigma_z \otimes \sigma_z$ representation is given by
\begin{equation}
    \ket{\psi}=a\ket{\uparrow\uparrow}+b\ket{\uparrow\downarrow}+c\ket{\downarrow\uparrow}+d\ket{\downarrow\downarrow}=\begin{pmatrix}
        a\\b\\c\\d
    \end{pmatrix},\; \text{with }|a|^2+|b|^2+|c|^2+|d|^2=1,\;a,b,c,d\in \mathbb{C}.\label{eq:ps}
\end{equation}
The eigen-states are frame dependent and $\{\ket{\lambda^L}\}$ corresponding the Eq.~\eqref{eq:eigenPL} with the chosen coordinate system in Fig.~\ref{fig:frame} are
\begin{align}
    &|\lambda_1^L\rangle=
    \begin{pmatrix}
        1\\ -\dfrac{ (1 + \alpha_{\psi}) \left( 2\alpha_{\psi}  \cos\theta +P_L (1-\alpha_{\psi} )\right)\sin\theta \, \mathrm{e}^{-\mathrm{i} \Delta\Phi}}{\sqrt{1 - \alpha_{\psi}^2} \left( \alpha_{\psi}\sin^2\theta  -  P_L (1 + \alpha_{\psi}) \cos\theta  + \sqrt{\alpha_{\psi}^2\left(1 + \cos^2\theta\right)^2  + P_L^2 \eta} \right)}\\ \dfrac{ (1 + \alpha_{\psi})\left(  2\alpha_{\psi}  \cos\theta -P_L (1-\alpha_{\psi} )\right)\sin\theta \, \mathrm{e}^{-\mathrm{i} \Delta\Phi}}{\sqrt{1 - \alpha_{\psi}^2} \left( \alpha_{\psi}\sin^2\theta  +  P_L (1 + \alpha_{\psi}) \cos\theta  + \sqrt{\alpha_{\psi}^2\left(1 + \cos^2\theta\right)^2  + P_L^2 \eta} \right)}\\1
    \end{pmatrix},\\
    &|\lambda_2^L\rangle=
    \begin{pmatrix}
       1\\ \dfrac{ (1 + \alpha_{\psi}) \left( 2\alpha_{\psi}  \cos\theta +P_L (1-\alpha_{\psi} )\right)\sin\theta \, \mathrm{e}^{-\mathrm{i} \Delta\Phi}}{\sqrt{1 - \alpha_{\psi}^2} \left( -\alpha_{\psi}\sin^2\theta  +  P_L (1 + \alpha_{\psi}) \cos\theta  + \sqrt{\alpha_{\psi}^2\left(1 + \cos^2\theta\right)^2  + P_L^2 \eta} \right)}\\ -\dfrac{(1 + \alpha_{\psi}) \left(  2\alpha_{\psi}  \cos\theta -P_L (1-\alpha_{\psi} )\right)\sin\theta \, \mathrm{e}^{-\mathrm{i} \Delta\Phi}}{\sqrt{1 - \alpha_{\psi}^2} \left( -\alpha_{\psi}\sin^2\theta  -  P_L (1 + \alpha_{\psi}) \cos\theta  + \sqrt{\alpha_{\psi}^2\left(1 + \cos^2\theta\right)^2  + P_L^2 \eta} \right)}\\1
    \end{pmatrix}.
\end{align}
Following the discussion at the end of Sec.~\ref{sec:bfc}, the eigen-states $\{\ket{\lambda^L}\}$ can be expanded within the spin-triplet subspace. Through swapping the $\hat{\mathbf{y}}$ and $\hat{\mathbf{z}}$ in Fig.~\ref{fig:frame}, we have
\begin{align}
\notag
    \ket{\lambda_1^L}&=\frac{P_L \left( (1 + \alpha_{\psi}) \cos^2\theta + (1 + \alpha_{\psi} \cos 2\theta) \right) + 2 \cos\theta \sqrt{\alpha_\psi^2 \sin^4\theta + P_L^2\eta}}{P_L (1 - \alpha_{\psi}) \sin^2\theta - 2 \alpha_{\psi} \cos\theta \sin^2\theta}\ket{1,1}+\ket{1,-1}\\
    & + \frac{\left(P_L \left( (1 + \alpha_\psi)^2 \cos^2\theta + \eta \right)+\sqrt{\alpha_\psi^2 \sin^4\theta + P_L^2\eta}\right)}{(1 + \alpha_\psi) \alpha_\psi \sin 2\theta - P_L \sin\theta (1 - \alpha_\psi^2)}\sqrt{2(1 - \alpha_\psi^2)} e^{i \Delta \Phi}\,\ket{1,0},\\
\notag
\ket{\lambda_2^L}&=\frac{P_L \left( (1 + \alpha_{\psi}) \cos^2\theta + (1 + \alpha_{\psi} \cos 2\theta) \right) - 2 \cos\theta \sqrt{\alpha_\psi^2 \sin^4\theta + P_L^2\eta}}{P_L (1 - \alpha_{\psi}) \sin^2\theta - 2 \alpha_{\psi} \cos\theta \sin^2\theta}\ket{1,1}+\ket{1,-1}\\
    & + \frac{\left(P_L \left( (1 + \alpha_\psi)^2 \cos^2\theta + \eta \right)-\sqrt{\alpha_\psi^2 \sin^4\theta + P_L^2\eta}\right)}{(1 + \alpha_\psi) \alpha_\psi \sin 2\theta - P_L \sin\theta (1 - \alpha_\psi^2)}\sqrt{2(1 - \alpha_\psi^2)} e^{i \Delta \Phi}\,\ket{1,0}.
\end{align}
Where $\ket{1, 1},\ket{1, 0},\ket{1, -1}$ are spin-triplet states:
\begin{equation}
\begin{aligned}
&\ket{1, 1} =\ket{ \uparrow\uparrow} \\
&\ket{1, 0} = \frac{1}{\sqrt{2}}(\ket{\uparrow\downarrow} + \ket{\downarrow\uparrow}) \\
&\ket{1, -1}= \ket{\downarrow\downarrow}
\end{aligned}
\end{equation}

\cleardoublepage
\bibliographystyle{jhep}
\bibliography{mainJHEP}

\end{document}